\documentclass[twocolumn,aps,prl,amsmath,superscriptaddress,notitlepage,longbibliography]{revtex4-1}
\usepackage{amssymb}
\usepackage{mathrsfs}
\usepackage{graphicx}
\usepackage{grffile}
\usepackage{float}
\usepackage[caption=false]{subfig}
\usepackage{scalerel}
\usepackage[pdftex,colorlinks=red]{hyperref}
\usepackage{amscd}

\usepackage{epstopdf}
\usepackage[normalem]{ulem}
\usepackage{verbatim}
\usepackage{xcolor,colortbl}

\usepackage{bm}
\usepackage{dcolumn}% Align table columns on decimal point

\usepackage[utf8x]{inputenc}

\definecolor{Gray}{gray}{0.9}

\def\blue{\textcolor{blue}}

\begin{document}

\def\qv{\vec{q}}
\def\blue{\textcolor{blue}}
\def\magenta{\textcolor{magenta}}
\def\apricot{\textcolor{Apricot}}

\definecolor{ora}{rgb}{1,0.45,0.2}
%{0.2, 0.7, 0.2}
\def\LH{\textcolor{black}}

\newcommand{\norm}[1]{\left\lVert#1\right\rVert}
\newcommand{\ad}[1]{\text{ad}_{S_{#1}(t)}}

%\title{Non-Hermitian skin effect for interacting particles on a flat band}
\title{Non-Hermitian skin clusters from strong interactions}
\author{Ruizhe Shen}
\affiliation{Department of Physics, National University of Singapore, Singapore 117542}
\author{Ching Hua Lee}
\email{phylch@nus.edu.sg}
\affiliation{Department of Physics, National University of Singapore, Singapore 117542}

%\pacs{73.43.Lp, 71.10.Pm}
\date{\today}

\begin{abstract}
	\section{abstract}
Strong, non-perturbative interactions often lead to new exciting physics, as epitomized by emergent anyons from the Fractional Quantum hall effect. Within the actively investigated domain of non-Hermitian physics, we provide a family of states known as non-Hermitian skin clusters. Taking distinct forms as Vertex, Topological, Interface, Extended and Localized skin clusters, they generically originate from asymmetric correlated hoppings on a lattice, in the strongly interacting limit with quenched single-body energetics. Distinct from non-Hermitian skin modes which accumulate at boundaries, our skin clusters are predominantly translation invariant particle clusters. As purely interacting phenomena, they fall outside the purview of generalized Brillouin zone analysis, although our effective lattice formulation provides alternative analytic and topological characterization. Non-Hermitian skin clusters originate from the fragmentation structure of the Hilbert space and may thus be of significant interest in modern many-body contexts such as the Eigenstate thermalization hypothesis (ETH) and quantum scars.
\end{abstract}

\maketitle
\section{Introduction}
When many-body interactions dominate single-particle energetics, unexpected physics often emerge~\cite{zhang1989effective,jain1989composite,jain1990theory,wen1992theory,zhang1995chern,stormer1999fractional,stormer1999nobel,ardonne2008degeneracy,ardonne2011structure,lee2018floquet,tournois2020braiding,tang2011high}. A classic example is the appearance of anyonic quasiparticles in Fractional Quantum hall (FQH) systems, whose emergent statistics cannot be inferred from single-particle Chern topology alone~\cite{tang2011high,regnault2011fractional,yao2013realizing,lee2014lattice,grushin2014floquet,yang2017generalized,abouelkomsan2020particle}. Indeed, the effects of strong non-perturbative interactions are determined by the structure~\cite{yang2020hilbert,sala2020ergodicity,patil2020hilbert,langlett2021hilbert,pietracaprina2021hilbert,lee2021frustration} of the many-body Hilbert space, not the first-quantized single particle description.
Lately, research have focused on non-Hermitian phenomena. Yet, the main avenues of non-Hermiticity - the non-Hermitian skin effect (NHSE) which in general leads to eigenstate accumulations along  boundaries~\cite{Lee2016nonH,gong2018topological,yao2018edge,lee2019anatomy, yokomizo2019non,lee2020unraveling,helbig2020generalized,schomerus2020nonreciprocal,li2021non,song2019non,borgnia2020non,zhang2020correspondence,okuma2020topological,kawabata2020higher,ghatak2020observation,xiao2020non,longhi2020non,mu2020emergent,lee2020ultrafast,zou2021observation,okuma2021quantum} and exceptional points from spectral singularities~\cite{berry2004physics,heiss2012physics,wu2014non,heiss2016circling,kozii2017non,leykam2017edge,hodaei2017enhanced,wang2019arbitrary,lafalce2019robust,okugawa2019topological,yoshida2019symmetry,yoshida2019exceptional,lee2020exceptional,park2020observation} - are essentially single-particle mechanisms based on first-quantized notions like non-Bloch eigenstates and single-particle band structure~\cite{yoshida2018non,yao2018non,lourencco2018kondo,yokomizo2019non,kawabata2020non,yokomizo2020non,nagai2020dmft,zhu2020photonic}. It is thus timely to ask if strong interactions can open the door to even more exotic phenomena. While several works have explored many-body effects and interactions in non-Hermitian settings~\cite{nakagawa2018non,hamazaki2019non,liu2020non,zhang2020skin,xi2021classification,zhai2020many,liu2020non,yoshida2021correlation}, interactions have not been the dominant physical mechanism.

In this work, we report the non-Hermitian skin clusters [Table~\ref{table1}] in the limit where interactions are much stronger than single-particle energetics. They are special translation invariant particle configurations, distinct from known non-Hermitian skin modes, being shaped by the connectivity structure of the many-body Hilbert space~\cite{yang2020hilbert,sala2020ergodicity,patil2020hilbert,lee2020many}, not the real-space lattice, and do not even require physical boundaries. They possess features like the loop gap rather than the existing line gap or point gap \cite{bessho2020topological}. Such interacting phenomena cannot be understood via the generalized Brillouin zone (GBZ), which has been highly successful in characterizing conventional skin modes~\cite{yao2018edge,lee2019anatomy, yokomizo2019non,lee2020unraveling,li2021non,li2020critical}.
%\footnote{But see Ref.~\cite{li2020critical} for exceptions.}
\section*{Results and discussion}
\begin{figure*}
	\includegraphics[width=.93\linewidth]{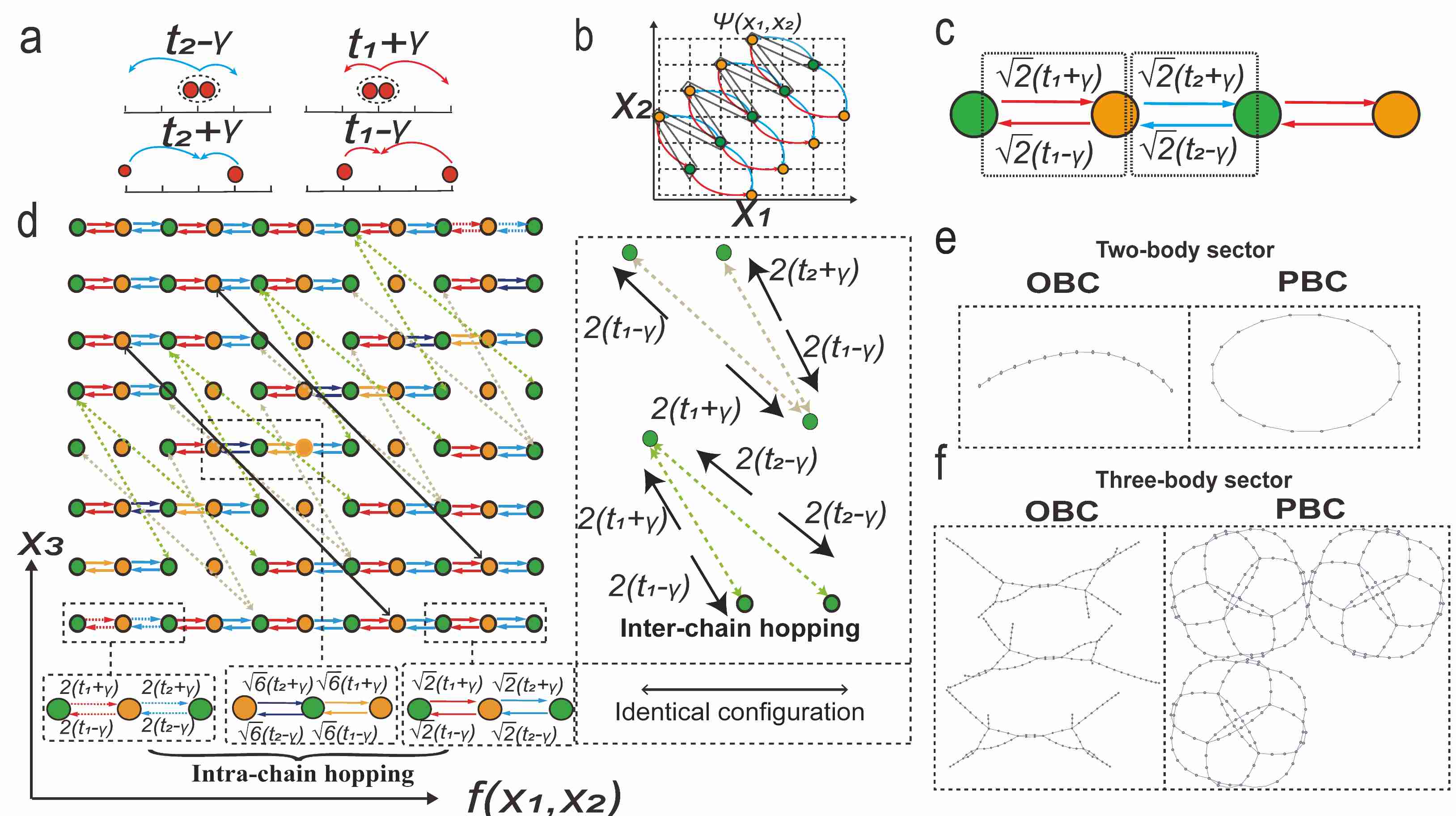}
	\caption{\textbf{ Sketch of effective models from correlated hoppings in Eq.~\eqref{H}.} \textbf{a} The four asymmetric two-body hoppings in our interacting model [Eq.~\ref{H}], non-Hermitian whenever $\gamma\neq 0$. \textbf{b} Two-boson configurations $(x_1,x_2)$ which are non-trivially coupled form a 1D subspace represented by a non-Hermitian effective Su–Schrieffer–Heeger (SSH) model \textbf{c} of  $2L-5$ (see methods), where $L$ is the number of sites in the physical OBC chain. \textbf{d} Effective 2D lattice for a 3-boson OBC subspace, which comprises an array of non-Hermitian SSH chains that are non-locally coupled by inter-chain hoppings (brown and green dashed double arrows), with certain physically identical configurations (black solid double arrows) ($(X(x_1,x_2),x_3)$ and permutations) identified. See Supplementary Note 1 for details of the effective 2D lattice. The labeled "intra-chain hoppling" is induced by bosonic statistics (see explanations in Supplementary Note 1). The ``identical configuration'' as the black double arrow is explained in the ``Methods'' section. \textbf{e-f} The Hilbert space connectivity graphs of $H$ in the 2 and 3-boson sectors for $L=15$, demonstrating non-trivial modifications to the graph structure from open boundary condition (OBC) to periodic boundary condition (PBC) beyond two bosons. 
	}\label{1}
\end{figure*}

\textbf{Emergent topology and non-locality in a minimal unbalanced two-body hopping model.} To understand how non-perturbative non-Hermitian interactions can lead to skin clusters states, we introduce a minimal 1D bosonic model purely consisting of unbalanced two-body correlated hoppings. Single-particle energetics are assumed to have been quenched, analogous to the scenario of dispersionless Landau levels~\cite{stone1992quantum,kapit2010exact,chung2010quasi,rhim2015landau,wang20173d,rhim2020quantum}. Our model contains the four simplest possible asymmetric two-body hoppings, such that one particle hops across one site and the other hops across two sites in the opposite direction:
[Fig.~\ref{1}\textbf{a}]:
\begin{equation}
\begin{aligned}
H=\sum_{i}^L(t_{1}+\gamma)c^{\dagger}_{i+2}c_{i}c^{\dagger}_{i-1}c_{i}
+(t_{1}-\gamma )c^{\dagger}_{i}c_{i+2}c^{\dagger}_{i}c_{i-1}\\
+(t_{2}-\gamma) c^{\dagger}_{i+1}c_{i}c^{\dagger}_{i-2}c_{i}
+(t_{2}+\gamma) c^{\dagger}_{i}c_{i+1}c^{\dagger}_{i}c_{i-2}
\end{aligned}
\label{H}
\end{equation}
where $\hat{c}_{i}$($\hat{c}_{i}^{\dagger}$) is the bosonic annihilation (creation) operator at the $i$-th site.  The non-Hermiticity is controlled by $\gamma$; when $\gamma=0$, each hopping process and its reverse occur with equal probability $t_1$ or $t_2$, depending on the direction of center-of-mass translation. Although this Hamiltonian may superficially resemble NHSE models with asymmetric single-particle hoppings i.e. the Hatano-Nelson model~\cite{gong2018topological,okuma2020topological}, it does not even act on single-particle states, implying that any skin cluster eigenstate must originate exclusively from particle-particle interactions.

\begin{table}
	\centering
	\begin{tabular}{|l|l|l|c|}\hline
		&\textbf{ Type of state}   & \textbf{ Location } & \textbf{OBC/PBC}\\ \hline
		1&Topological edge mode & Physical boundary & OBC \\ \hline
		2&Skin edge mode & Physical boundary & OBC \\ \hline
		
		%{\it 3}&{\it Vertex skin cluster} & Intersection & OBC \\ \hline
		%
		%{\it 4}&{\it Topological skin cluster}&Effective interface&OBC\\ \hline
		%
		%{\it 5}&{\it Interface skin cluster} & Effective interface & PBC \\ \hline
		%
		%{\it 6}&{\it Extended skin cluster} & Effective interface & PBC \\ \hline
		%
		%{\it 7}&{\it Localized skin cluster} & Non-local region & PBC \\ \hline
		{3}&{Vertex skin cluster} & Intersection & OBC \\ \hline
		
		{4}&{Topological skin cluster}&Effective interface&OBC\\ \hline
		
		{5}&{Interface skin cluster} & Effective interface & PBC \\ \hline
		
		{6}&{Extended skin cluster} & Effective interface & PBC \\ \hline
		
		{7}&{Localized skin cluster} & Non-local region & PBC \\ \hline
	\end{tabular}
	\caption{\textbf{Characterization of non-Hermitian skin clusters.} The 7 types of boundary states of our two-boson hopping model $H$ in Eq.\eqref{H}, with the 5 types of unconventional states (3-7). Vertex (3) and Topological (4) skin clusters require both effective and physical boundaries [Fig.~\ref{2}\textbf{a} for open boundary conditions (OBCs)]. Interface (5), Extended (6), and Localized (7) skin clusters are purely due to the effective interface in $H$ [Fig.~\ref{3}\textbf{a} for periodic boundary conditions (PBCs)]. Only (1) and (2) have been reported in the literature. The states (3), (4) and (5) are specific to 3-boson clusters, while states (6) and (7) can exist in generic $N>2$-boson clusters (see Supplementary Note 1 for more details).}
	\label{table1}
\end{table}
The many-body Hilbert space of our model can be systematically dissected by identifying its various subsectors and how they are coupled~\cite{hudomal2020quantum}. We start by observing that each two-boson sector harbors hidden $\mathbb{Z}$ topology associated with 1D chiral symmetry. To concretely see that, note that two bosons only interact if they are either on the same site or three sites apart [Fig.~\ref{1}\textbf{a}]. Hence, by restricting to only such configurations [Fig.~\ref{1}\textbf{b}], we can index the 2-boson subspace as an effective \emph{single}-body 1D chain [Fig.~\ref{1}\textbf{c}] in configuration space. To see that, we take a two-boson state, $\psi(x_{1},x_{2})$ which is identified with $\psi(x_2,x_1)$ due to bosonic asymmetry, located at sites~ $x_1$ and $x_2\geq x_1$, and introduce a relabeling $\Psi(X(x_{1},x_{2}))$ for the same state, as given by
\begin{equation}
X(x_{1},x_{2})=
\left\{\begin{array}{rcl}
2i-3 & \text{ if } & x_{2}=x_{1}=i, \\
2i-2 & \text{ if } & x_{2}=i+2,x_{1}=i-1
\end{array}\right. 
\label{eq2}
\end{equation}
that captures configurations $(x_1,x_2)$ acted by $H$. All other configurations decouple trivially. As illustrated in Fig.~\ref{1}\textbf{c} and elaborated in the ``Methods'' section, the 2-boson sector reduces to an effective 1D non-Hermitian Su–Schrieffer–Heeger (SSH) chain with asymmetric couplings proportional to $t_1\pm \gamma$ and $t_2\pm \gamma$, since odd sites physically represent double bosonic occupancy and even sites represent bosons separated by three physical lattice spacings. We emphasize that generic models with asymmetric multi-particle hoppings often also contain variants of generalized SSH models~\cite{lieu2018topological,yao2018edge,marques2018topological}.

The merits of our abovementioned basis relabeling become evident when we consider three or more particles. Key features of the many-body Hilbert space structure already emerge in a 3-boson sector. We write a three-boson state as $\psi(x_1,x_2,x_3)=\Psi(X(x_{1},x_{2}),x_{3})$, with permuted particle coordinates understood to correspond to the same state. A 3-boson sector then becomes a collection of coupled non-Hermitian SSH chains indexed by $x_3$ [Fig.~\ref{1}\textbf{b}]. Their effective couplings come in two types: (i) dashed asymmetric couplings representing physical 2-body couplings between $x_3$ and $x_1$ or $x_2$, weighted by boson degeneracy factors such as $\sqrt{2!}, \sqrt{2!^2}$ and $\sqrt{3!}$, and (ii) solid double arrows representing pairs of identified sites that correspond to the same physical site due to bosonic symmetry i.e. $(X(x_1,x_2),x_3)$ and $(X(x_2,x_3),x_1)$ for $x_2=x_1+3=x_3-3$. Also, due to bosonic symmetry, some asymmetric couplings become effectively non-local in the resultant lattice. For instance, consider three bosons, two close together and acted by a local physical interaction, and the third far away from both. But due to symmetry under particle exchange, we have an equivalent description where the faraway boson is swapped with one of the nearby bosons, such that the local interaction appears as a non-local effective interaction.

\textbf{Skin cluster states and Hilbert space fragmentation.} Since interactions are reduced to ordinary hoppings on the effective lattice that represents the physically interesting many-body sector, they can be understood in terms of single-particle concepts. We show in Fig.~\ref{1}\textbf{e} and Fig.~\ref{1}\textbf{f} the Hilbert space connectivity under open boundary conditions (OBCs) and periodic boundary conditions (PBCs), for 2-boson and 3-boson subspaces respectively (see Supplementary Note 2 for that of 4 and 5-boson subspaces). While the 2-body subspaces simply form 1D chains, as previously explained, the 3-body spaces decouple into three sectors, each containing some states that are not easily reached, suggestive of Hilbert space fragmentation~\cite{yang2020hilbert,pietracaprina2019hilbert,sala2020ergodicity,lee2020frustration,patil2020hilbert,langlett2021hilbert,herviou2021many}. For more than two particles, the Hilbert space connectivity graphs become much more complicated when going from OBCs to PBCs, much beyond merely ``closing up'' chains into loops. This hence hints of more exotic phenomena besides the conventional NHSE. Indeed, in the PBC case,  there exists a hierarchy of smaller loops that are coupled only at a few selected states. When superimposed onto an effective lattice structure, these inter-loop couplings correspond to the previously mentioned effective non-local hoppings.

\begin{figure}
	\includegraphics[width=.99\linewidth]{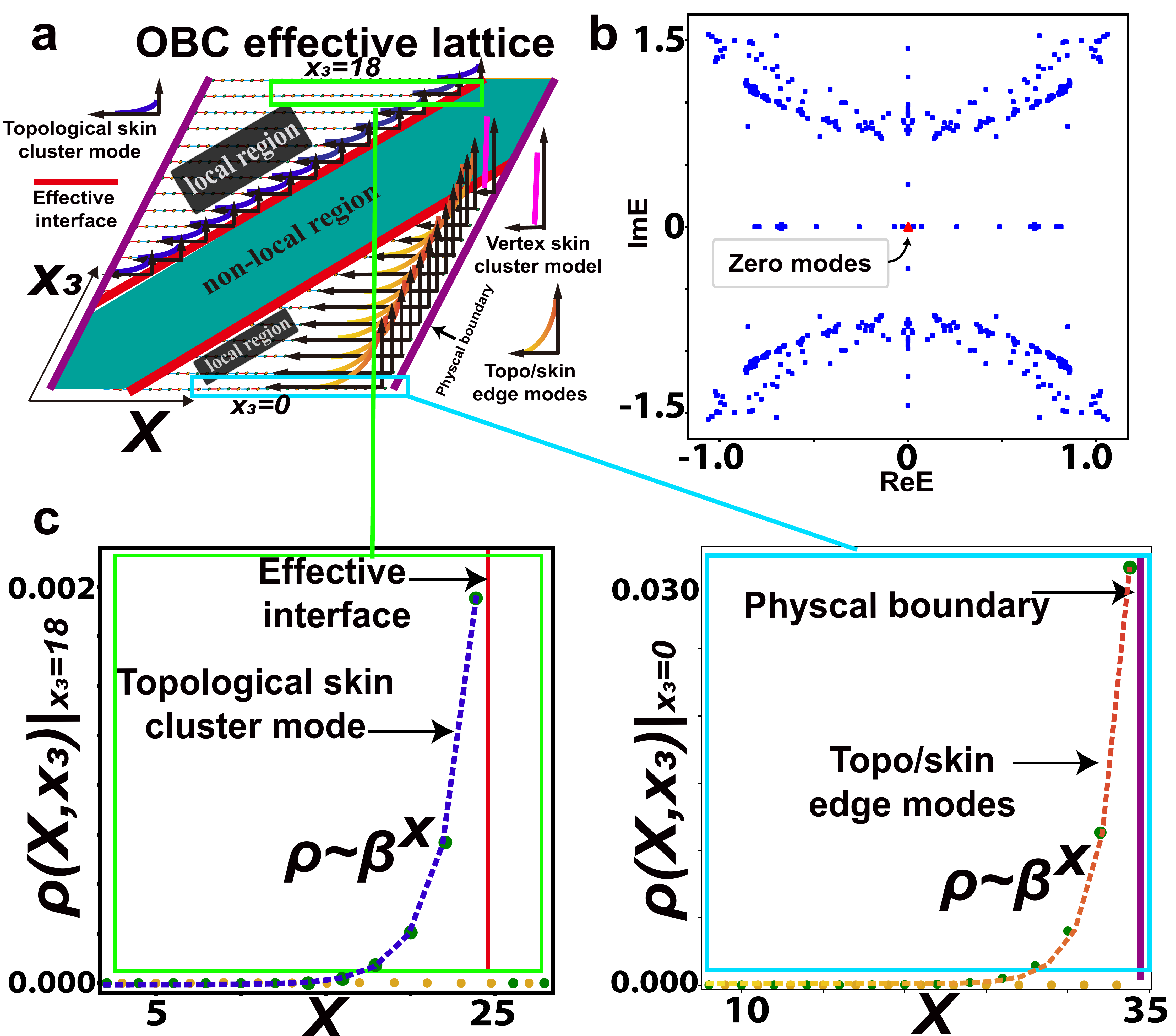}
	\caption{ \textbf{Localizations in the effective lattice for the 3-boson sector under open boundary conditions (OBCs).} \textbf{a} Schematic of the effective Hilbert space lattice for the 3-boson sector under OBCs. Regions with local (white) and non-local (cyan) effective hoppings are demarcated by the interaction-induced effective interfaces (red lines). Topological skin cluster modes accumulate against effective interfaces, while vertex skin clusters are localized at intersections between effective interfaces and physical edges (purple), which also host ordinary skin/topo edge modes. \textbf{b} The three-boson OBC spectrum with parameters $t_{1}=1.0,t_{2}=0.2,\gamma=0.8$, on a $(2L-5)\times L$ effective lattice ($L=20$). Zero modes correspond to either topological edge modes or skin clusters. \textbf{c} Spatial density profiles $\rho(X,x_3)=\left| \Psi(X,x_3)\right|^{2}$ of both types of topological modes, which clearly accumulate at either the effective interface or physical boundary with identical decay profiles $\rho(X,x_3)\sim \beta^X$, $\beta=(t_{1}+\gamma)/(\gamma-t_{2})$ (see methods). Green and yellow dots in \textbf{c} correspond to the horizontal even and odd  sites in \textbf{a}. More details of the exponential localizations can be found in in Supplementary Note 1.}
	\label{2} 
\end{figure}

\begin{figure*}
	\includegraphics[width=.7\linewidth]{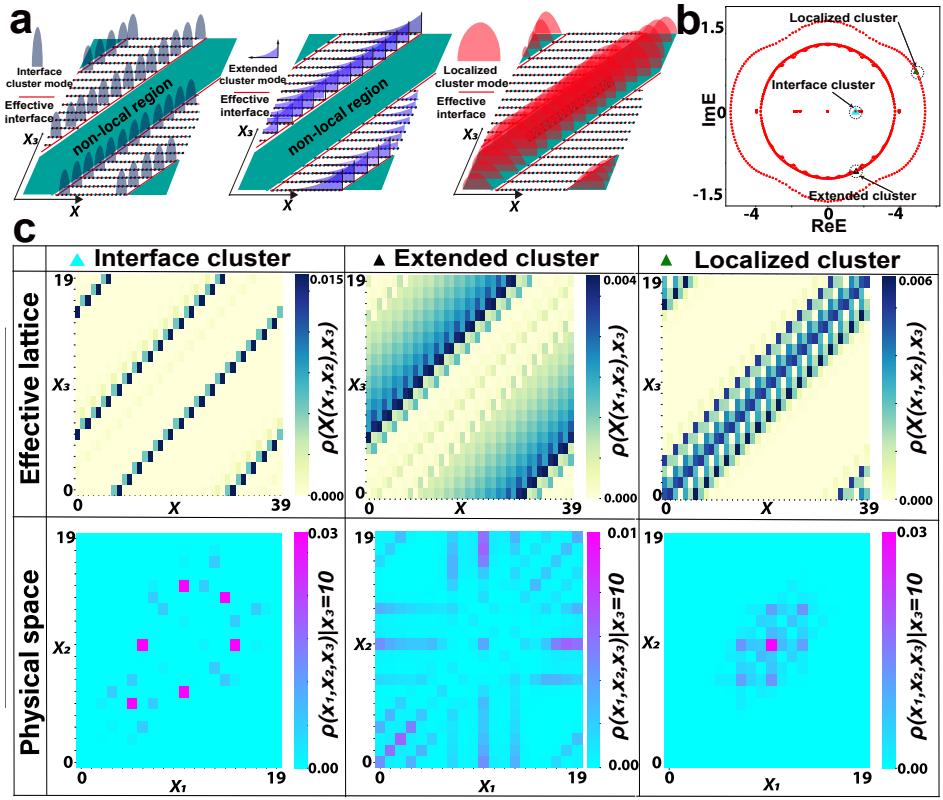}
	\caption{\textbf{Characterization of different clusters in the $N=3$-boson sector under periodic boundary conditions (PBCs).} \textbf{a} Schematics of the effective Hilbert space lattice for the 3-boson sector under PBCs, with each panel illustrating its respective type of PBC cluster mode. The effective interfaces in the red lines generate a ``non-local region" in the lattice. Interface cluster modes (grey) and extended cluster modes (blue) appear at the boundary of the non-local region, whereas localized cluster modes exist in the bulk. \textbf{b} The three-boson PBC spectrum with parameters $t_{1}=1.4,t_{2}=1.2,\gamma=0.5,L=20$. Generally, localized cluster states form an outer loop with the largest $|E|$, separated from the inner loop of Extended skin clusters via a loop gap. Exact solutions for interface clusters form isolated spectra are given in Supplementary Note 1. \textbf{c} Numerically computed spatial density distribution of the three types of clusters. $\rho(X,x_3)=|\Psi(X(x_{1},x_{2}),x_{3})|^{2}$ represents the density in the effective lattice, and $\rho=|\psi(x_{1},x_{2},x_{3})|^{2}_{x_{3}=10}$ is for the density in the physical lattice. The color bars in \textbf{c} indicate the amplitude of each type of density. Interface clusters, while seemingly constrained at the effective interface, actually consist of bosons several sites from each other. Extended clusters favor physical configurations with two overlapping bosons and another far away. Localized clusters contain bosons that are physically localized within sites from each other.}
	\label{3}
\end{figure*}

For concrete analysis of our specific model, we decompose the effective lattice of Fig.~\ref{1}\textbf{c} into regions that take different roles in forming the skin clusters [Fig.~\ref{2}\textbf{a} for OBCs,~\ref{3}\textbf{c} for PBCs]. We shall focus on clusters with $N=3$ bosons, and then discuss how some of them generalizes to generic $N$ number of bosons. There exists a ``non-local'' region around $x_3\approx X(x_1,x_2)$ which is crossed by a large number of non-local effective hoppings across several effective sites (diagonal effective hoppings in Fig.~\ref{1}\textbf{c}, which hop across the cyan ``non-local'' region in Fig.~\ref{2}\textbf{a} or Fig.~\ref{3}\textbf{a}). They arise from physical interactions between a two-boson sector and a third boson. The background ``local'' region experiences only nearest-neighbor SSH-type effective hoppings (white in Fig.~\ref{2}\textbf{a} or \ref{3}\textbf{a}) from 2-boson processes only. Note that all effective hoppings originate from the one and only type of interaction term present in $H$; the demarcation into ``local'' and ``non-local'' regions reflect the qualitatively different roles of processes involving different numbers of bosons. Non-local hoppings across the ``non-local'' region destroy the periodicity in the effective lattice, precluding any GBZ description that relates the distinct OBC and PBC spectra (Fig.~\ref{2}\textbf{b},~\ref{3}\textbf{b}), unlike in non-interacting models.

Importantly, effective interfaces emerge between the ``local'' and ``non-local'' regions, in addition to physical boundaries at the edges of the effective lattice, if any. This results in additional localization behavior beyond topological and skin edge localizations. We identified seven distinct types of boundary states [Table \ref{table1}]. 
As effective interfaces are induced by particle statistics, not physical boundaries, our states (Types (3) to (7)) are named ``skin cluster states''. Vertex skin clusters (3) are unique highly-localized states at point intersections (vertices) of effective and physical boundaries [Fig.~\ref{2}\textbf{a}], and are elaborated in Supplementary Note 1. Topological (4), Interface (5), Extended (6), and Localized (7) skin clusters represent distinct ways by which the effective interfaces exert topological and skin localization and will be elaborated below.

The five types of states from (3) to (7) in Table \ref{table1} are distinguished by the location at which they are localized in the effective space. Under OBCs, there exist ``topological skin clusters" (see the blue mode in FIG.\ref{2}) which are mathematically equivalent to usual topological boundary modes at the physical boundaries. At the crossing point of the physical and effective boundary, the localization is further squeezed into highly localized ``vertex skin clusters'' represented as the pink mode in FIG.\ref{2}. When we consider PBCs, where the physical boundary is no longer present, we obtain more types of states without conventional OBC analogs.  Instead of ``topological skin clusters'', we now have ``interface skin clusters", ``extended skin clusters" and ``localized skin clusters",  which are represented by gray, blue, and red modes in FIG.\ref{3}. As suggested by their name, ``interface skin clusters" are highly localized at the effective boundaries, while the other two types, ``localized skin clusters" and ``extended skin clusters", appear in the coupled and noncoupled chains in the effective lattice. Among these five states, only the ``localized skin clusters" and ``extended skin clusters" have direct generalized in the higher dimensional Hilbert space of $N>3$-boson clusters ($N>3$-boson sector). We will further elaborate on these states in FIG.\ref{5} below and surrounding paragraphs, as well as in Supplementary Note 1.

Although skin clusters also arise from asymmetric couplings, they differ from conventional NHSE skin states in a few important ways: (i) they only exist when two or more particles interact; (ii) they exist due to the inhomogeneity of effective couplings from particle statistics, \emph{not} physical boundaries as in the NHSE~\cite{lee2020many}: in fact (5-7) exist under PBCs, not OBCs; (iii) while the NHSE usually gives rise to non-local responses~\cite{helbig2020generalized,pan2020non,lee2020unraveling}, here the skin clusters themselves are already the consequence of non-local effective couplings; (iv) due to the ``non-local'' region, the various types of skin clusters do not possess GBZ descriptions, unlike ordinary skin states.

\textbf{Topological skin clusters.} Topological skin clusters eigenstates (4) are topological zero modes at the effective interface between the ``local'' and ``non-local'' regions. Physically, they are manifestations of how the inherent SSH-type topology in a 2-boson sector interplays with the presence of a third boson. Although effective interfaces exist identically across both OBC [Fig.~\ref{2}] and PBC [Fig.~\ref{3}] settings, Topological skin clusters only exist under OBCs. PBC skin clusters (5-7) differ significantly from them in both energetics [Figs.~\ref{2}\textbf{b} vs.~\ref{3}\textbf{b}] and spatial profile [Figs.~\ref{2}\textbf{c} vs.~\ref{3}\textbf{c}]. This enigma arises because of the non-local effect of local unbalanced hoppings: Under PBCs, the absence of physical boundaries (edges) leads to significant interference between the NHSE-like accumulation from either side of the non-local region, destroying ``unadulterated'' topological states. That topological skin clusters (4) mathematically result from the same topological mechanism as ordinary topological edge states (1) can be seen from their vanishing energies [Fig.~\ref{2}\textbf{b}], and identical localization lengths [Fig.~\ref{2}\textbf{c}] of $(\log\beta)^{-1}$, with $\beta=(t_1+\gamma)/(\gamma-t_2)$ as is well-known for the non-Hermitian SSH model~\cite{yao2018edge}. See ``Methods'' for edge states in the non-Hermitian SSH model.

\textbf{Translation invariant skin clusters.} Without physical boundaries (PBCs), the three types of skin clusters (5-7) that exist are of special non-topological origins. They are Interface (5), Extended (6) and Localized (7) skin cluster, so-called because they respectively exist at the interface between the ``local'' and ``non-local'' regions (5), are extended across the wide ``local'' region (6) or are localized within the relatively narrow ``non-local'' regions (7), as seen from $|\Psi(X(x_{1},x_{2}),x_{3})|^{2}$ density plots of Fig.~\ref{3}\textbf{c}. Although Interface clusters (2) appear the most localized of the three on the effective lattice, in physical coordinates, their density plot $|\psi(x_{1},x_{2},x_{3})|^{2}$ (Fig.~\ref{3}\textbf{c}) reveal strong correlations between particles a few sites apart. By contrast, Localized skin clusters (7) consist of 3 bosons that are indeed almost physically overlapping, ``caged'' by the non-local effective hoppings. The extended skin clusters (6) are the most delocalized but are still considered clusters states because they are exponentially localized along the effective interface. 

The eigenenergies of different types of skin cluster eigenstates generically form very distinct loci in the complex energy plane. As shown in Fig.~\ref{3}\textbf{b}, extended and localized cluster states (6,7) form two concentric spectral loops, while the interface states (5) form isolated points in the loop interiors. Under parameter tuning, these characteristic spectral loci remain largely robust, even though they may distort and cross each other. 
 \begin{figure*}
 	\includegraphics[width=\linewidth]{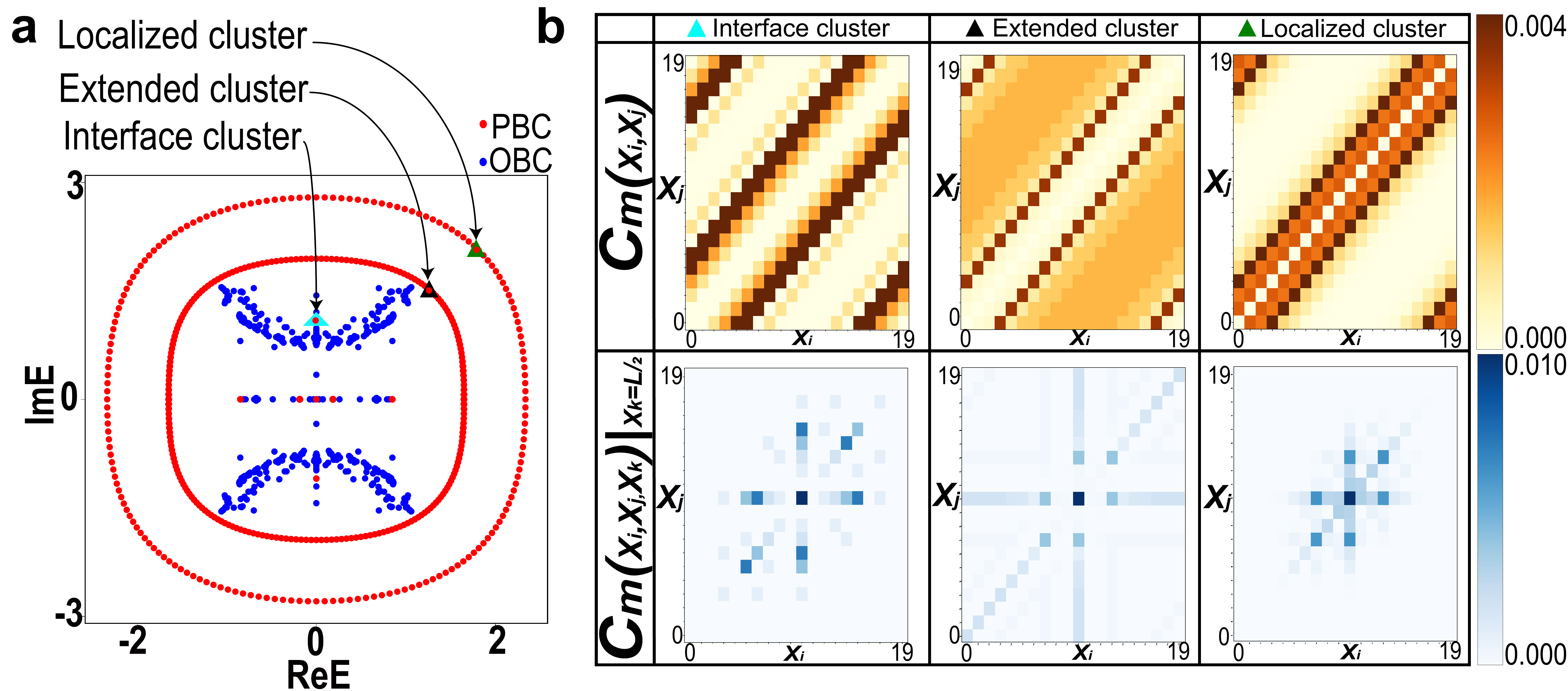}
 	\caption{\textbf{Correlation behaviors for different clusters in the $3$-boson sector under periodic boundary conditions (PBCs).} \textbf{a} The 3-boson PBC spectrum (red) is characterized by a loop gap between concentric loops, which also enclose isolated PBC interface cluster energies and most of the OBC spectrum (blue). The selected localized, extended, and interface cluster are labeled in green, black, and blue triangular dots respectively. \textbf{b} Density correlators for representative PBC eigenstates $\psi_m$ of the three cluster types from \textbf{a}. 2-site correlators $C_{m}(x_{i},x_{j})$ in Eq.~\eqref{co}~$(a)$ for $i\neq j$ reveal that interface configurations $|x_i-x_j|=3$ are favored in all cluster types. 
 		3-site $C_{m}(x_{i},x_{j},x_{k})|_{x_k=L/2}$ in Eq.~\eqref{co}~$(b)$ is evaluated at $x_k=L/2$. Without loss of generality, due to translation symmetry, reveal that extended clusters are most likely to have a doubly occupied site, but that localized clusters possess the shortest correlation range. Parameters are $t_{1}=1.0,t_{2}=0.2,\gamma=0.8$, with $L=20$. The color bars in \textbf{b} represent the amplitude of each correlator respectively.
 	}
 	\label{4}
 \end{figure*} 
\noindent{\textit{Structure of skin cluster states and the loop gap.--}} 
The concentric rings of eigenenergies in the 3-body PBC spectrum [Fig.~\ref{3}\textbf{b}] are robust across a large range of parameters, and can even persist even beyond three bosons, as elaborated in Supplementary Note 1. We name this unmistakable separation of concentric PBC spectral loops as a \emph{loop gap}, in the spirit of point gaps for loops which encircle interior points~\cite{okuma2020topological}. To understand why the loop gap occurs, we note that the outer/inner spectral ring corresponds to localized/extended cluster states.  
In the fully localized limit where all bosons move rigidly like a single composite particle, the PBC spectrum traces out a large loop given by $E=\sum_j t_j e^{ipj}$, where $p\in[0,2\pi)$ and $t_j$ are the effective $j$-site hoppings acting on the composite particle. Compared to localized clusters, extended clusters must trace out smaller loops because at least one boson does not move in unison with the others [Fig.~\ref{3}\textbf{c}], leading to ``destructive interference''. Note that localized and extended clusters move freely as a whole, unlike interface clusters and topological edge modes which are constrained to specific physical locations. This causes the latter two to be spectrally enclosed within the localized and extended clusters [Fig.~\ref{4}\textbf{a}] since, being confined, they cannot be significantly amplified by the asymmetric hoppings and must thus possess energies close to $E=0$.

The above arguments can be corroborated by the following 2 and 3-site density correlators, which reveal more insight into the real-space structure of the cluster states:
\begin{equation}\label{co}
\begin{aligned}
&\,(a)~C_{m}(x_{i},x_{j})=\left|\left\langle \psi_{m} \left| n_in_j \right|\psi_{m} \right\rangle  \right|,(i\neq j)
\\
&\,(b)~C_{m}(x_{i},x_{j},x_{k})=\left|\left\langle \psi_{m} \left| n_{i}n_{j}n_{k} \right|\psi_{m} \right\rangle  \right|,
\end{aligned}
\end{equation}
where $\psi_{m}$ is the selected eigenstate and the density operator $n_i=c^{\dagger}_{i}c_{i}$ measures the occupancy at site $i$. These correlators are plotted in Fig.~\ref{4}\textbf{b} for representative eigenstates $\psi_m$ as specified in Fig.~\ref{4}\textbf{a}; to put their magnitudes in perspective, note that they scale respectively as $L^{-2}$ and $L^{-3}$ for a perfectly uniform state, but remain at unity for a perfectly clustered state. Comparing the 2-site correlator $C_m(x_i,x_j)$ in Fig.~\ref{4}\textbf{b} with the single-site density plots in Fig.~\ref{3}\textbf{c}, both for the effective lattice, we observe similar trends despite the different model parameters and chosen states. However, $C_m(x_i,x_j)$ always reveals conspicuously strong spectral weights at the interfaces, suggesting that the bosons prefer to be separated by 3-5 sites regardless of the cluster type. The three site-correlator $C_{m}(x_{i},x_{j},x_{k})$ further reveals that extended states tend to consist of 2 bosons on the same site, together with another arbitrarily far boson.

\textbf{Effective interface and clustering in many-body sectors.} Many results from the three-boson sector, which we elaborated above, continue to hold in sectors i.e. clusters with arbitrarily number of bosons. In essence, effective boundaries in the $N$-body Hilbert space continue to play a crucial role, and are generalized to be regions of high connectivity that separate regions of different levels of connectivities. In this context, what is required is that there exists a contrast between the connectivities of the nodes in the many-body Hilbert space, rather than particular values of the connectivities. This is because the non-Hermitian skin cluster effect acts whenever there exists inhomogeneities in the Hilbert space graph, reminiscent of the usual non-Hermitian skin effect which are non-perturbatively sensitive to any form of spatial inhomogeneity.

Shown in FIG.\ref{5} are the Hilbert space graphs of $N=3,4$ and $5$ bosons, as well as illustrative localized clusters and delocalized states. The graph in FIG.\ref{5} \textbf{a} is equivalent to the two-dimensional lattice in FIG.\ref{3} \textbf{a}. For the three-boson sector, we already know that the effective interface separates the graph into different connectivity regions (see FIG.\ref{3}). Similarly, for the four and five-boson sectors in FIG.\ref{5} \textbf{e},\textbf{f} and \textbf{i},\textbf{j}, analogous effective interfaces exist at regions of highest connectivities, such that they also separate regions of different connectivities (see FIG.\ref{5} \textbf{d}, \textbf{h}, \textbf{l}, dashed blue boxes). Generally, a node with more neighbors indicates that the corresponding state is a more strongly correlated state. Thus, the effective interface represents kinetic blockage constraints due to interactions(FIG.\ref{5} \textbf{d}). Such a notion of an effective boundary can be directly generalized to arbitrary number of particles. From FIG.\ref{5} \textbf{c}, \textbf{g}, and \textbf{k}, we also see that localized skin cluster states always exist on the outer spectral loop regardless of the number of particles, while other illustrative delocalized states exist within the loop gap. Around an effective interface, the states take the form of a cluster of particles that accumulate against each other, hence the name ``skin cluster states''. 
\begin{figure*}
	\centering
	\includegraphics[width=0.5\linewidth]{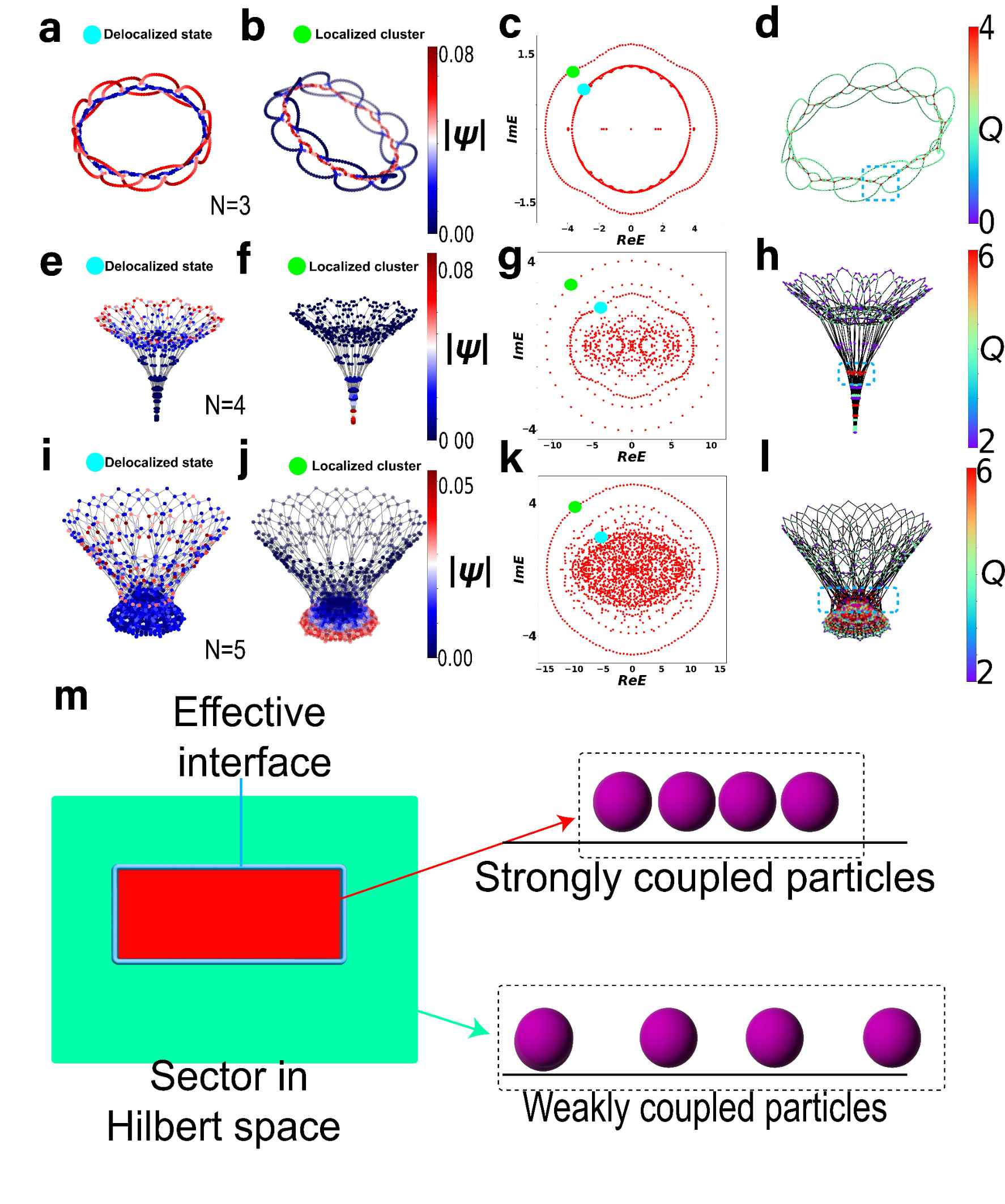}
	\caption{
		\textbf{Many-body Hilbert space graphs and the densities of selected cluster and delocalized non-cluster states in $N=3,4,5$-boson sectors under periodic boundary conditions (PBCs).} Plot \textbf{a} represents the delocalized state (state indicated by the light blue dot in \textbf{c}), and plot \textbf{b} depicts the localized cluster state (state indicated by the green dot in \textbf{c}).
		Plot \textbf{a} and \textbf{b} share the same color bar which represents the amplitude of the wavefunction $|\psi|$ at each node. \textbf{c} is for the spectra under PBCs with the selected states for \textbf{a} and \textbf{b}. In plot \textbf{d}, the effective boundary highlighted in blue dashed box is the red region of higher connectivity (no. of graph neighbors $Q$ identified in the color bar) that separates regions of relatively low connectivity. The representations for \textbf{e}-\textbf{h} and \textbf{i}-\textbf{l} follow the manner of the plots \textbf{a}-\textbf{d}. According to \textbf{a}-\textbf{l}, we find that localized cluster states always lie on the outer spectral loop regardless of the number of bosons $N$. Also, they always lie next to effective boundaries, which are regions of highest connectivity on the Hilbert space graph that also separates regions of relatively high and low connectivity.
		\textbf{m} Schematic notion of an "effective interface" in a Hilbert space graph: by separating regions of different levels of connectivity, it is physically interpreted as a demarcation between strongly and weakly coupled particles. The size and filling are \textbf{a}-\textbf{d} L=20, N=3; \textbf{e}-\textbf{h} L=10, N=4; \textbf{i}-\textbf{l} L=10, N=5. Parameters are $t_{1}=1.4,t_{2}=1.2,\gamma=0.5$. }
	\label{5}
\end{figure*}

\section{Conclusion}
 Our simple model with asymmetric two-body hoppings (Eq.~\eqref{H}) is a paradigmatic representation of non-reciprocal non-Hermitian physics in the strongly interacting limit where single-body energetics are quenched. While asymmetric couplings have been associated with the well-known single-particle NHSE, our purely interacting model exhibits various clustering behavior [Table~\ref{table1} (3-7)]. As a departure from conventional NHSE, these cluster states are mostly translation invariant, even if they require OBCs, as in topological skin clusters (4). 

Stemming from effective interfaces in a fragmented Hilbert space, our states should exist in generic non-Hermitian interacting lattices with (i) asymmetric couplings and (ii) quenched single-particle energetics. Unlike in Hermitian models where the effective interfaces are physically significant only if they scale with the system size, our non-Hermitian effective interfaces can dramatically modify the system dynamics even if they are of vanishing fractal dimension in the Hilbert space, thanks to the profoundly non-perturbative and non-local nature of non-Hermitian skin clustering.  As such, the Hilbert space fragmentation from non-Hermitian skin clustering, which violates the eigenstate thermalization hypothesis (ETH)~\cite{sala2020ergodicity,pietracaprina2019hilbert,lee2020frustration,rakovszky2020statistical}, will likely inspire further study into the non-equilibrium dynamics in non-Hermitian systems, e.g., quantum many-body scars~\cite{choi2019emergent,pietracaprina2019hilbert,zhao2020quantum,turner2018weak,turner2018quantum,turner2020correspondence}, and many-body localization \cite{pal2010many,nandkishore2014many,alet2018many}. Appropriately generalized, our model may be physically realized with density-assisted tunneling in driven optical lattices~\cite{zhao2019engineered,zhao2020quantum}, quantum digital computer circuits with suitably arranged imaginary time evolution~\cite{mcardle2019variational,lin2021real}, 
or, topolectrical circuits~\cite{lee2018topolectrical,helbig2020generalized,olekhno2020topological,lee2020imaging, liu2020octupole,yang2020observation, wang2020circuit,zou2021observation,stegmaier2021topological} at the effective lattice level.
\section{Methods}
\textbf{Two-boson sector as a non-Hermitian SSH chain.} In order to deal with our non-Hermitian interacting model in Eq.~\eqref{H}, we apply its Hilbert space graph, which serves as an effective non-interacting model. In this section, we use a minimal two-body model for Eq.~\eqref{H} to explain how the NHSE works in the Hilbert space graph. According to the analysis in the main text, the Hilbert space graph of the two-body model is the effective SSH chain. Here we furnish more details of this effective SSH chain, which indirectly controls the physics of the two bosons in physical space.

\begin{figure}
	\includegraphics[width=.99\linewidth]{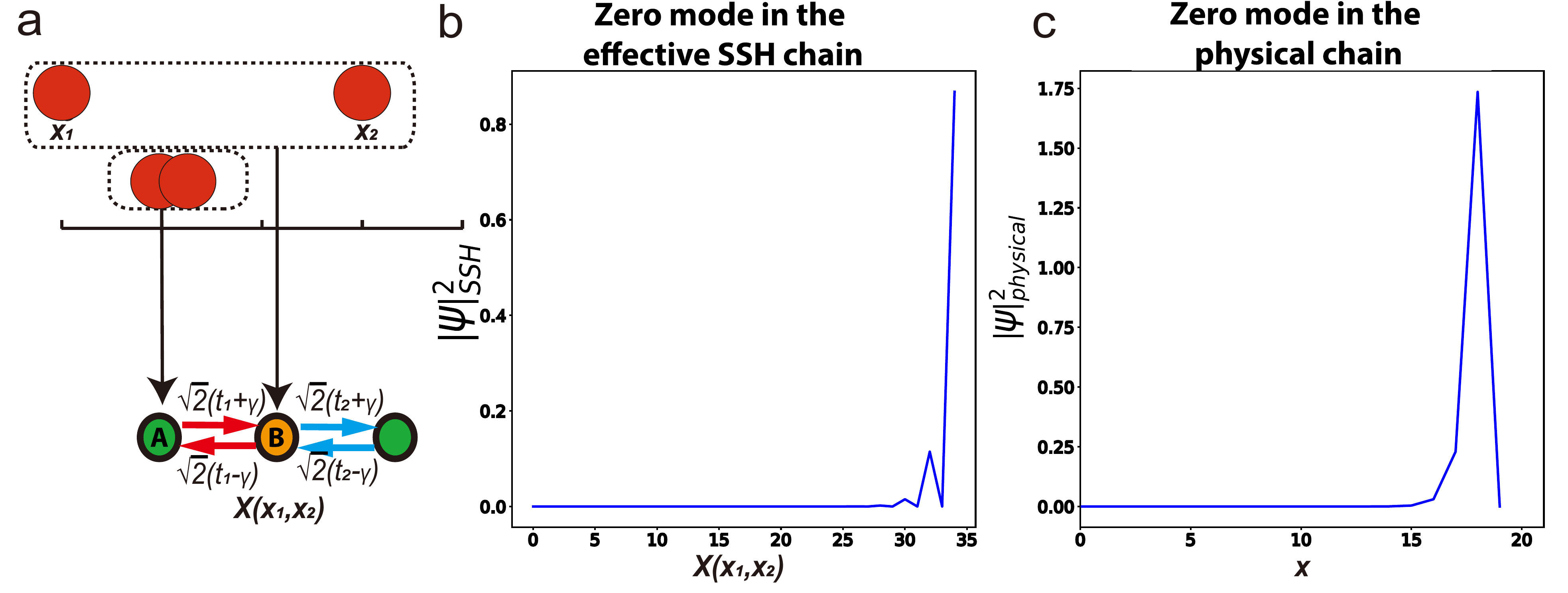}
	\caption{\textbf{Correspondence of localizations in the real space and the effective chain for the two-boson sector under open boundary conditions (OBCs).} \textbf{a} The  Non-Hermitian effective SSH model corresponding to the physical two-boson model, and the corresponding two-boson state from the physical chain that maps to odd/even sites in the effective SSH chain. The squared amplitudes of the same zero mode in the effective chain \textbf{b} vs. the physical chain \textbf{c} with $t_{1}=1.2,t_{2}=0.2,\gamma=1.0,L=20$. The corresponding length of the effective SSH chain is $2L-5$. $X(_{1},x_{2})$ in \textbf{b} is given in Eq.~\ref{eq2}, and $x$ in \textbf{c} represents the position in the physical space.}
	\label{6}
\end{figure}

We consider the zero mode in the effective SSH chain, expressed as $\psi =(\psi_{1A},\psi_{1B},......~\psi_{n-1A},\psi_{n-1B},\psi_{nA})^{T}$ (A,B are the labels of the sublattices in Fig.~\ref{6}). The sublattice wave function  takes the exponential form $(\psi_{nA},\psi_{nB})_{T}=\beta^{n}(\psi_{1A},\psi_{1B})^{T}$ according to the generalized Brillouin zone ansatz~\cite{yao2018edge, lee2019anatomy}. In the bulk, the eigen-equation satisfies 
\begin{equation}
	\begin{aligned}
		\sqrt{2}(t_{1}-\gamma)\psi_{iB}+\sqrt{2}(t_{2}+\gamma)\psi_{i-1B}=E\psi_{iA}
		\\\sqrt{2}(t_{1}+\gamma)\psi_{iA}+\sqrt{2}(t_{2}-\gamma)\psi_{i+1A}=E\psi_{iB}
	\end{aligned}
\end{equation}
At the boundary, the OBC boundary condition gives
\begin{equation}
	\begin{aligned}
		\sqrt{2}(t_{1}-\gamma)\psi_{1B}=E\psi_{1A}\\
		\sqrt{2}(t_{2}+\gamma)\psi_{n-1B}=E\psi_{nA}.
	\end{aligned}
\end{equation}
From them, we can obtain the solution to the zero mode of SSH chain as $\psi_{1B}=0,\beta=(t_{1}+\gamma)/(\gamma-t_{2})$. Fig.~\ref{6}\textbf{b} presents a zero mode in the effective SSH chain while Fig.~\ref{6}\textbf{c} presents its corresponding squared amplitude in physical space, as a two-boson system. Similarly, the three-boson sector can be mapped to a two-dimensional non-Hermitian model as shown in FIG.~\ref{1}\textbf{d}. See more details of the effective two-dimensional non-Hermitian model in Supplementary Note 1.
\begin{figure}
	\includegraphics[width=1\linewidth]{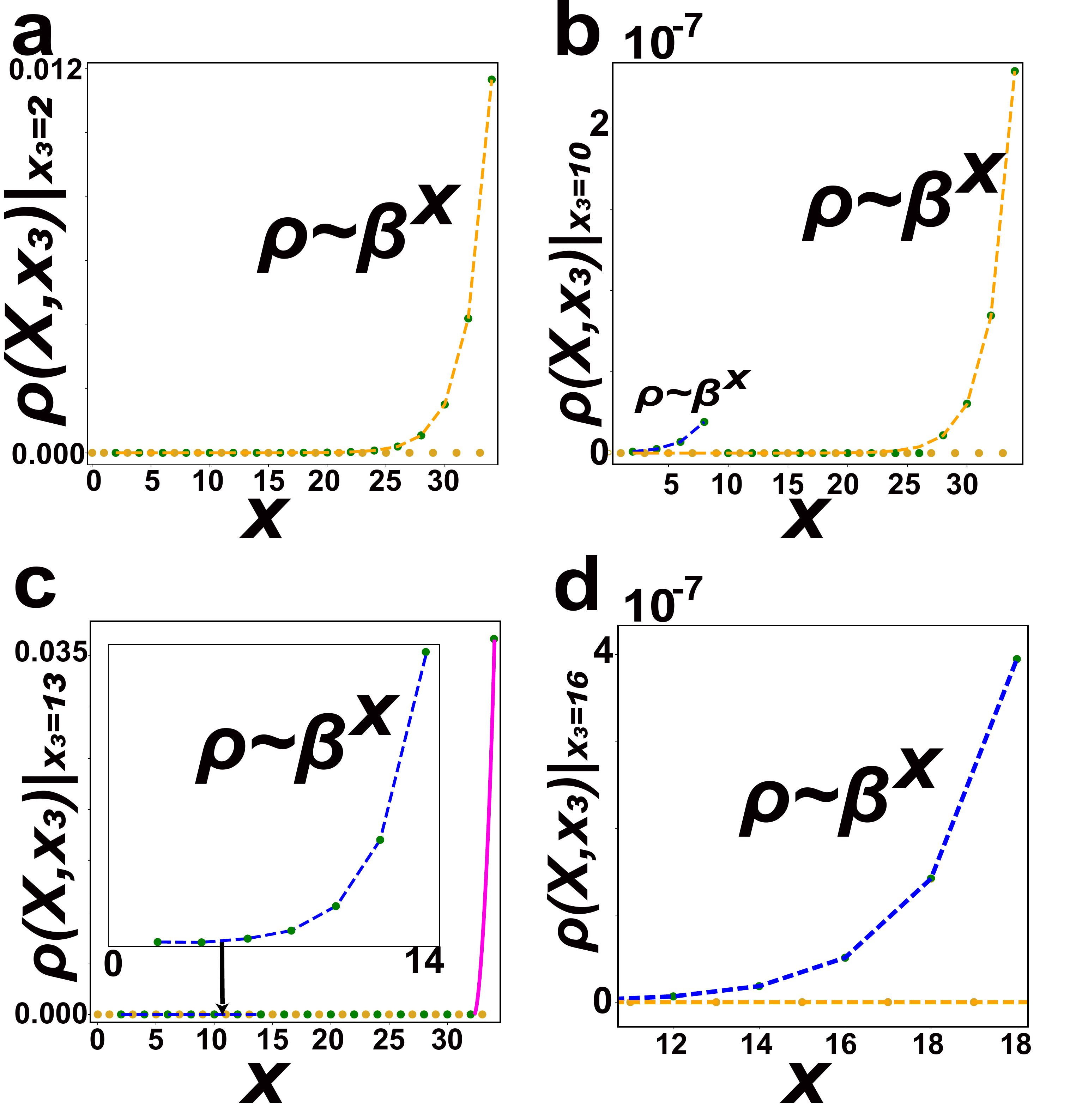}
	\caption{\textbf{Exponential  localizations in the effective lattice for 3-boson cluster under open boundary conditions (OBCs).} 2D cross-sectional density $\rho=\left| \Psi(X,x_{3})\right|^{2}$ of the OBC effective lattice in FIG.\ref{1}\textbf{d}, with fixed $x_{3}=2$ \textbf{a}, $10$ \textbf{b}, $13$ \textbf{c}, $16$ \textbf{d} of selected states with $t_{1}=1.0,t_{2}=0.2,\gamma=2.0,L=20$. For the cases $x_{3}=2,10$, there is skin localization at the physical boundary(the yellow curves). For the cases $x_{3}=10,13,16$, there is skin localization at the interface (the blue curves). Also present are the trivial mode and the vertex localization (the pink curve in \textbf{c}) at the intersection between the physical boundary and the interface in the effective lattice. All the blue and yellow curves exhibit the exponential localizations as $\rho\sim\beta^{x}$ with $\beta=(t_{1}+\gamma)/(\gamma-t_{2})$. }
	\label{7}
\end{figure} 
\textbf{Cluster states of topological origin}. In our effective lattice for 3 bosons under OBCs from the main text, there exists both physical boundaries and effective interfaces as shown in FIG.\ref{2}\textbf{a}. When $\gamma \neq 0$, localization occurs at both the two boundaries.  Fig.~\ref{7} presents the state amplitude distribution in the effective lattice of a selected OBC zero mode with $t_{1}=1.2,t_{2}=0.2,\gamma=0.5,L=20$. 

To show that the localizations near the interface and the physical boundary have the same topological origin, we perform the curve fitting of these decays and compare the analytical result $\beta=(t_{1}+\gamma)/(\gamma-t_{2})$ (the solution of the effective non-Hermitian SSH model) with the curve fitted results from numerical diagonalization of the 3-body system. Indeed, for the zero modes of the three-body sector, the decays at the physical boundary ($\beta^{\prime}$) and effective interfaces ($\beta^{\prime\prime}$) all agree viz. $\beta^{\prime\prime},\beta^{\prime}\approx \beta$ ($\rho(x)\approx \beta^{x}$).

\begin{figure*}
	\centering
	\includegraphics[width=0.9\linewidth]{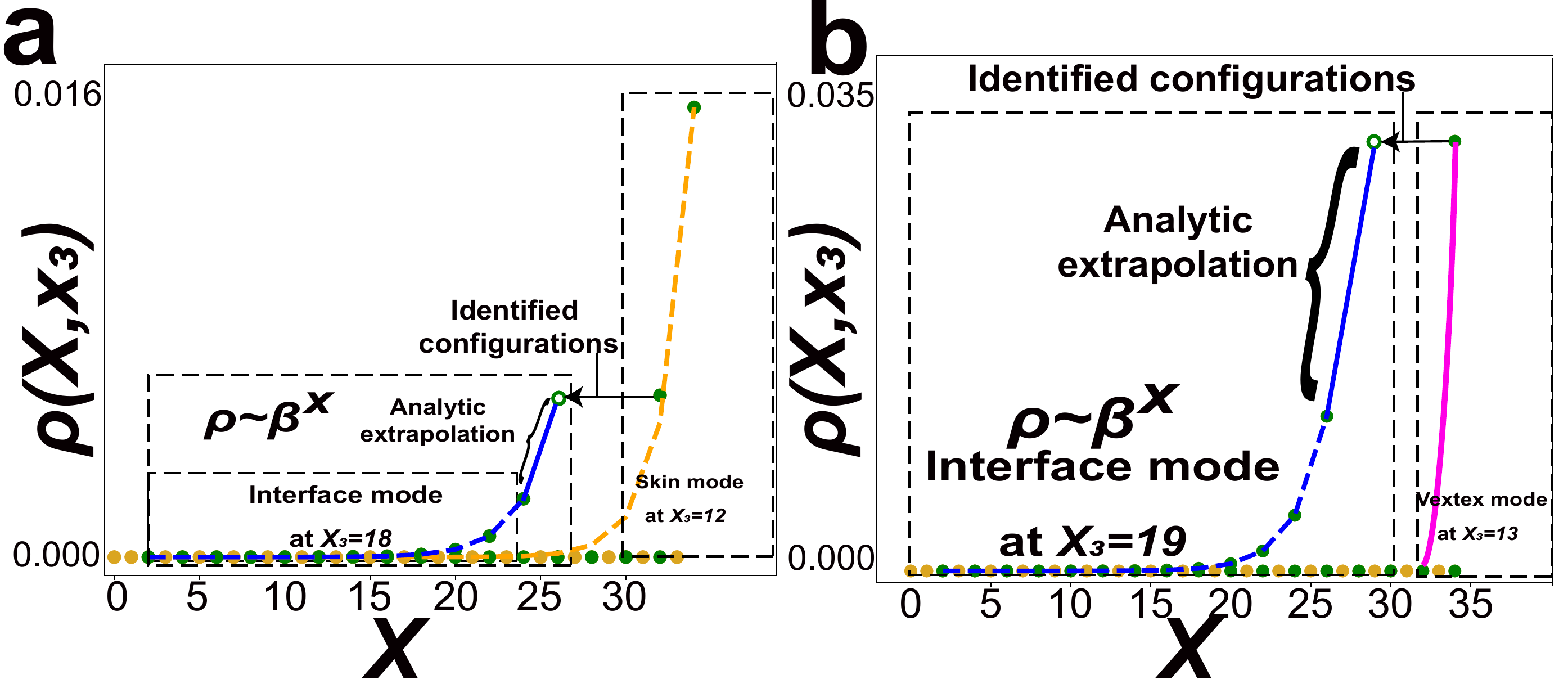}
	\caption{\textbf{Identified configurations from bosonic symmetry}. \textbf{a} The yellow curve: the skin mode of $\rho(X,x_{3})=\left| \Psi(X,x_{3})\right|^{2}_{x_{3}=12}$; The blue curve: the interface  mode of $\rho(X,x_{3})=\left| \Psi(X,x_{3})\right|^{2}_{x_{3}=18}$. The ``Identified configurations'' between sites allows us to move a section (the dashed rectangle) of the skin mode to the nearest site of the interface mode (the green hollow circle). We perform an analytic extrapolation (blue solid line) according to the non-Bloch scaling controlled by $\beta$ (the solid blue line) to link the two ends. \textbf{b} The pink curve: the vertex mode of $\rho(X,x_{3})=\left| \Psi(X,x_{3})\right|^{2}_{x_{3}=13}$; The blue curve: the interface  mode of $\rho(X,x_{3})=\left| \Psi(X,x_{3})\right|^{2}_{x_{3}=19}$. Exponential localizations as the blue and yellow curves follow the scaling $\rho(X,x_{3})\sim\beta^{X}$ with $\beta=(t_{1}+\gamma)/(\gamma-t_{2})$. In our model, bosonic symmetry leads to equivalent ``identified configurations'' in the effective lattice in FIG.~\ref{1}\textbf{d}, and hence skin cluster states on it that can be pieced together like a jigsaw-puzzle. The density as $\rho(X,x_{3})=\left| \Psi(X,x_{3})\right|^{2}$ represents the localization in the graph in FIG.~\ref{2}\textbf{a} for the 3-boson cluster under open boundary conditions (OBCs). Parameters for \textbf{a} and \textbf{b} are $t_{1}=1.0,t_{2}=0.2,\gamma=2.0$, and $L=20$.}
	\label{8}
\end{figure*}

The interface within the effective lattice is demarcated by ``identified configurations'' arising from bosonic statistics (See the structure of the 2D lattice in Supplementary Note 1). So as to explain these interface localizations under OBC, we plot skin (vertex) modes and the interface mode connected by the "Identified configurations" together (Fig.~\ref{8}). These results suggest that the interface mode is a branch of the skin (vertex) mode, such that we can use analytic extrapolation to attach the interface mode to a section of the skin (vertex) mode.

\section{Data availability}
The data that support the plots within this paper are available upon reasonable request.
\section{Code availability}
Computer codes used to generate the plots in this work are available upon reasonable request via email to C.H.L.
\section{Acknowledgements}
We thank Hui Jiang and Yin Zhong for discussions. The Hamiltonians are numerically computed with QuSpin\cite{weinberg2017quspin,weinberg2019quspin}, and R.S. thank Phillip Weinberg for providing suggestions on the computational programming. This work is supported by the MOE Tier I start-up grant WBS: R-144-000-435-133.

\section{Author contributions}
C.H.L. carried out preliminary studies and supervised the project. R.S. carried out additional theoretical and computational studies. Both authors discussed the results and participated in the writing of the manuscript.
\section{Additional information}
\textbf{Competing Interests:} The authors declare no competing interests.
\section{References}
\bibliography{references}
\newpage

\setcounter{equation}{0}
\setcounter{figure}{0}
\setcounter{table}{0}
\setcounter{section}{0}
\renewcommand{\theequation}{S\arabic{equation}}
\renewcommand{\thefigure}{S\arabic{figure}}
\renewcommand{\thesection}{S\arabic{section}}
\onecolumngrid
\flushbottom
\newpage
%\newpage

\appendix

\section{Supplementary Information for ``Non-Hermitian skin clusters from strong interactions"}

The supplementary materials are organized according to the following sections:\\
\\
1. Details of the effective lattice construction for our model, its topological properties, and its Hilbert space connectivity structure.\\
2. Details of the effective boundaries resulting in various types of 3,5 and 5-boson clusters, including analytic characterizations and alternative measures of clustering, as well as the scaling properties of their Hilbert subspaces.

%1. The details and the analytical solution to the two-boson model
%
%2. The verification of the effective model
%
%3. The details and the illustration on boundary state in the three-boson model
%
%4. The details and the explanation for correlation function and the excitation gap

\section{Supplementary Note 1: The effective lattice - construction details and connectivity structure}\label{Note1}
Our interacting model comprises exclusively of asymmetric two-boson correlated hoppings introduced in Eq.~1 of the main text:
\begin{equation}
	\begin{aligned}
		H=\sum_{i}^L(t_{1}+\gamma)c^{\dagger}_{i+2}c_{i}c^{\dagger}_{i-1}c_{i}
		+(t_{1}-\gamma )c^{\dagger}_{i}c_{i+2}c^{\dagger}_{i}c_{i-1}\\
		+(t_{2}-\gamma) c^{\dagger}_{i+1}c_{i}c^{\dagger}_{i-2}c_{i}
		+(t_{2}+\gamma) c^{\dagger}_{i}c_{i+1}c^{\dagger}_{i}c_{i-2}
	\end{aligned}
	\label{H}
\end{equation}

As a result, non-trivial physics requires at least two bosons, unlike usual single-particle non-Hermitian skin effect (NHSE) models \cite{yao2018edge, lee2019anatomy}. The main physics hosted by this interacting model appears whenever we have at least 3 particles. Below, we hence elaborate on its connectivity structure in the 2-boson and 3-boson sectors of its Hilbert space.

\subsection{Three-boson sector}

\begin{figure}[h]
	\centering
	\includegraphics[width=\linewidth]{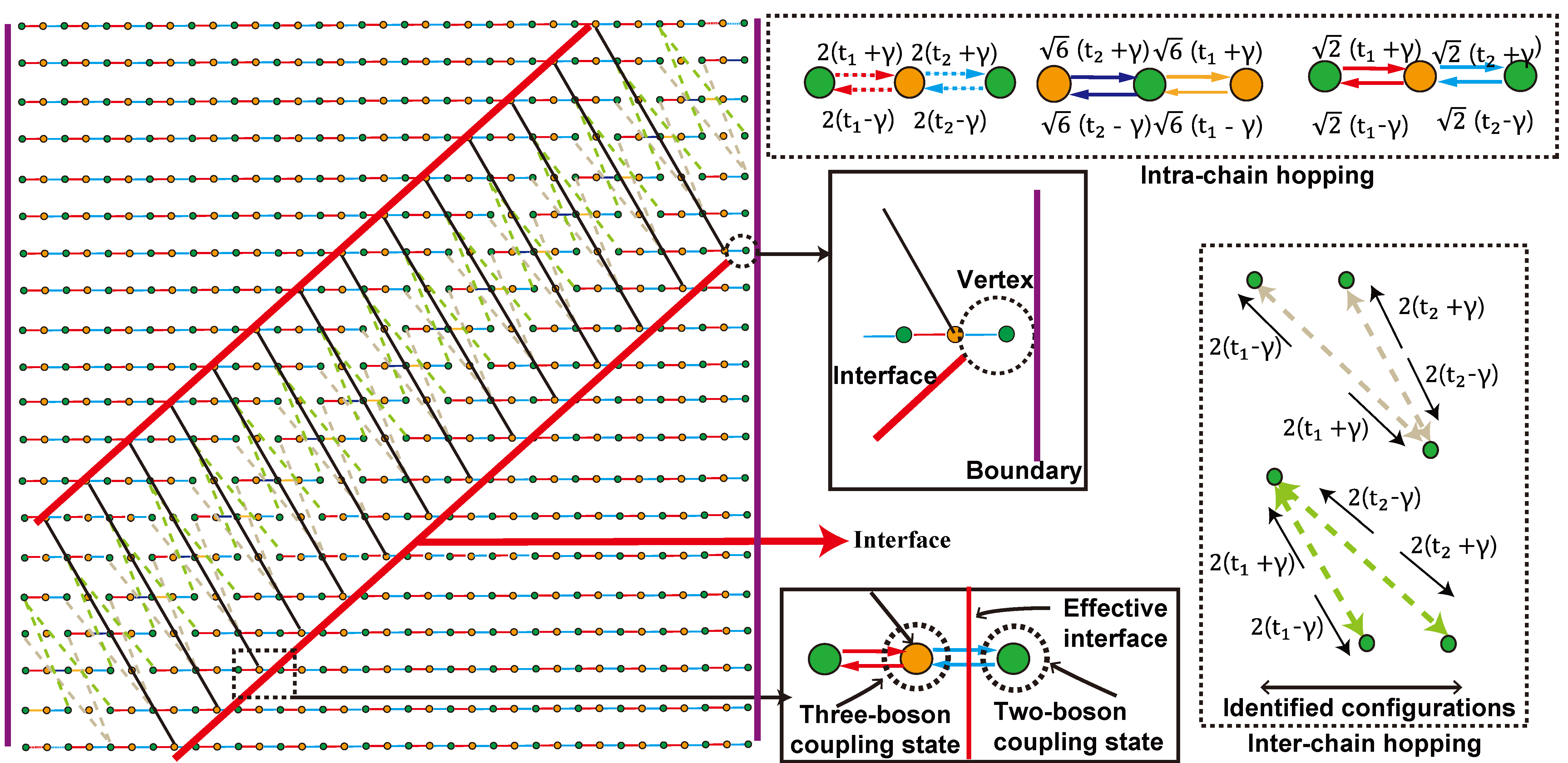}
	\caption{Schematic of the OBC effective model for three bosons, shown for an effective lattice of size $(2L-5)\times L$ $(L=20)$. The combined effects of the boson-boson couplings and bosonic statistics give rise to the inter-chain hoppings. The effective interface appears between the central ``non-local region'' between the two red diagonal lines with many inter-chain hoppings, and the ``local region'' with only intra-chain hoppings. In the non-local region, the inter-chain hoppings depend on the positions of all three bosons, unlike the local region where the effective Su–Schrieffer–Heeger (SSH) hoppings depend only on the positions of two of the bosons. The states in green dots represent the state with $\left|x,x,x^{\prime} \right\rangle $, and the states in yellow dots represent the state with $\left|x-1,x+2,x^{\prime} \right\rangle $.} 
	\label{fig:suppfig1}
\end{figure}

We now construct the effective lattice for the 3-boson sector of our model $H$ (see Fig.~\ref{fig:suppfig1} for the effective lattice under OBCs). In the following, we explain the details of mapping a state from the 3-boson Hilbert space to the effective lattice. We express the three-boson state in the physical Hilbert space as $\left| x_{1},x_{2},x_{3}\right\rangle$. According to the transformation Eq.~2 of the main text:

\begin{equation}
	X(x_{1},x_{2})=
	\left\{\begin{array}{rcl}
		2i-3 & \text{ if } & x_{2}=x_{1}=i, \\
		2i-2 & \text{ if } & x_{2}=i+2,x_{1}=i-1
	\end{array}\right. 
	\label{eq2}
\end{equation}
so that the three-boson state can be written on the effective lattice as
\begin{equation}
	\begin{aligned}
		\left| x_{1},x_{2},x_{3}\right\rangle=\left| X(x_{1},x_{2}),x_{3}\right\rangle
	\end{aligned}
\end{equation}

As a result, the two-body hoppings in the physical space becomes effective hoppings in the effective 2D lattice. In Fig.~\ref{fig:suppfig1}, there are two types of hoppings in the effective lattice. The intra-chain hoppings (A1)(A2)(A3) come from the two-body hoppings (if the boundary is present, $\left|1,X(1,4)\right\rangle \neq \left|X(1,1),4)\right\rangle$)

\begin{equation}
	\begin{aligned}
		&(A1):\qquad c^{\dagger}_{1}c_{2}c^{\dagger}_{4}c_{2} \left|1,X(2,2)\right\rangle=2\left|1,1,4\right\rangle=2\left|1,X(1,4)\right\rangle\\
		&(A2):\qquad c^{\dagger}_{i-1}c_{i}c^{\dagger}_{i+2}c_{i} \left|X({i},{i}),{i}\right\rangle=\sqrt{6}\left|X({i-1},{i+2}),{i}\right\rangle\\
		&(A3):\qquad c^{\dagger}_{i-1}c_{i}c^{\dagger}_{i+2}c_{i} \left|X({i},{i}),{j}\right\rangle=\sqrt{2}\left|X({i-1},{i+2}),{j}\right\rangle, \qquad (|{j}-{i}|>3)\\
	\end{aligned}
\end{equation}
The inter-chain hoppings in FIG.\ref{fig:suppfig1} come from
\begin{equation}
	\begin{aligned}
		&(1)~c^{\dagger}_{i-1}c_{i}c^{\dagger}_{i+2}c_{i} \left|X({i},{i}),{i+2}\right\rangle
		=2\left|{i-1},{i+2},{i+2}\right\rangle=2\left|X({i+2},{i+2}),{i-1}\right\rangle\\
		&(2)~c^{\dagger}_{i-1}c_{i}c^{\dagger}_{i+2}c_{i} \left|X({i},{i}),{i-1}\right\rangle
		=2\left|{i-1},{i+2},{i-1}\right\rangle=2\left|X({i-1},{i-1}),{i+2}\right\rangle\\
		%&(3)~c^{\dagger}_{i-2}c_{i}c^{\dagger}_{i+1}c_{i} \left|X({i},{i}),{i-2}\right\rangle
		%=2\left|{i+1},{i-2},{i-2}\right\rangle=2\left|X({i-2},{i-2}),{i+1}\right\rangle\\
		&(3)~c^{\dagger}_{i+1}c_{i+3}c^{\dagger}_{i+1}c_{i} \left|X({i+3},{i+3}),{i}\right\rangle
		=2\left|{i+1},{i+1},{i+3}\right\rangle=2\left|X({i+1},{i+1}),{i+3}\right\rangle\\
	\end{aligned}
\end{equation}
There exists equivalent labels of states like $\left| X({i-3},{i}),{i+3}\right\rangle=\left| {i-3},X({i},{i+3})\right\rangle
$, which are indicated as "Identified configurations" in Fig.~\ref{fig:suppfig1}.

\begin{figure}[h]
	\centering
	\includegraphics[width=1\linewidth]{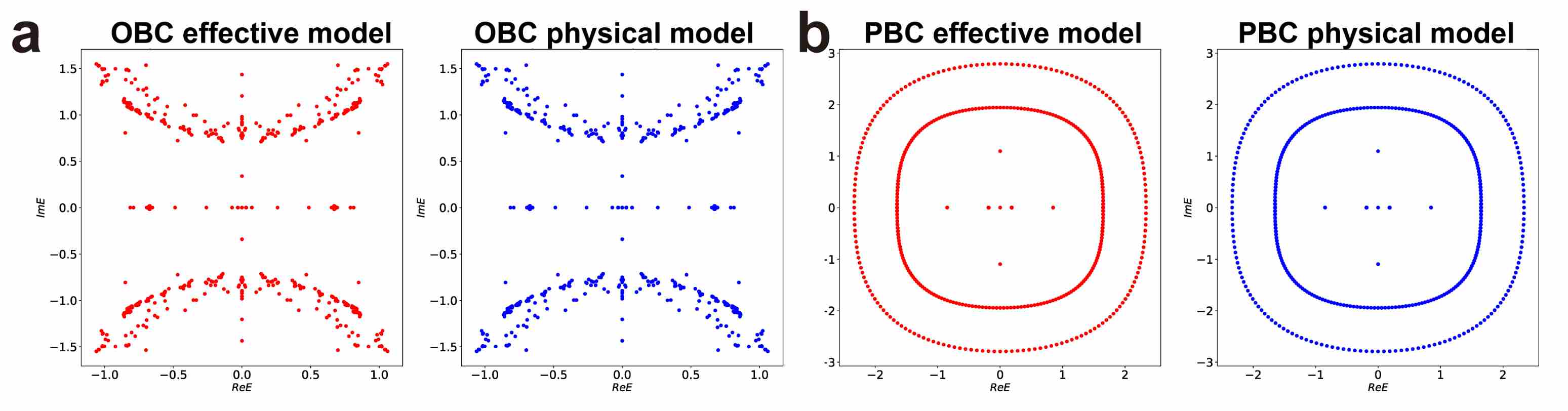}
	\caption{Comparison of the 3-boson OBC and PBC spectra between our physical (red) and effective (blue) models, with exact agreement as demonstrated. Parameters are $t_{1}=1,t_{2}=0.2,\gamma=0.8,L=20$.  }
	\label{fig:suppfig2}
\end{figure}

To check the validity of our effective model, we compare the PBC and OBC spectrum of the physical and effective models in FIG.\ref{fig:suppfig2}, and the spectrum results are all in agreement.

Additional to the main text, we list the 5 new states with bosonic clusters in Table I induced by "effective interfaces" in FIG.\ref{fig:suppfig1}. 
\begin{table}[h]
	\centering
	\begin{tabular}{|l|l|l|c|c|}\hline
		&\textbf{ Type of state}   & \textbf{ Location } & \textbf{OBC/PBC} & \textbf{Example of states}\\ \hline
		%		1&Topological edge mode & Physical boundary & OBC &\\ \hline
		%		2&Skin edge mode & Physical boundary & OBC & \\ \hline
		%		\rowcolor{Gray}
		1&{\it Vertex skin cluster} & Intersection & OBC & $\left|L-1,L-1,L-6 \right\rangle $\\ \hline
		2&{\it Topological skin cluster}&Effective interface&OBC& $\left|x-2,x,x \right\rangle (2<x<L-1) $\\ \hline
		3&{\it Interface skin cluster} & Effective interface & PBC& $\left|x-2,x,x+2 \right\rangle $\\ \hline
		4&{\it Extended skin cluster} & Effective interface & PBC& $\left|x-2,x,x+d \right\rangle (d>2) $\\ \hline
		5&{\it Localized skin cluster} & Non-local region & PBC& $\left|x,x,x \right\rangle $\\ \hline
	\end{tabular}
	\caption{The 5 new types of states in our two-boson hopping model present in 3-boson clusters, and some representative examples.}
	\label{tables1}
\end{table}

According to the examples in Table~\ref{tables1}, we categorize the states in term of the distance between the particles.
In FIG.\ref{fig:suppfig1}, for the horizontal chain, the vertex skin cluster appears at the critical region where the physical boundary and effective interface crosses:

\begin{equation}
	\left|L-3,L,L-6 \right\rangle-{\rm Effective~interface}-\left|L-1,L-1,L-6 \right\rangle-{\rm Boundary}
\end{equation}
where ``-'' denote couplings across many-body states in the Hilbert space graph.

The extended skin cluster, interface skin cluster, localized skin cluster appear at the effective interface according to the following configurations:

\begin{itemize}
	\item
	extended skin cluster: 
	\begin{equation}\left|x-2,x,x+d \right\rangle ...-{\rm Effective~interface};
	\end{equation}
	\item interface skin cluster: 
	\begin{equation}\left|x-2,x,x+2 \right\rangle-{\rm Effective~interface};
	\end{equation}\\
	\item localized skin cluster: 
	\begin{equation}-{\rm Effective~interface}-...\left|x,x,x \right\rangle ...-{\rm Effective~interface}.
	\end{equation}.\\
\end{itemize}
For these 5 new states in our three-boson sector, the ``extended skin cluster'' and ``localized skin cluster'' can be generalized to systems and clusters with more particles (the manner in which they are localized in higher-dimensional Hilbert spaces are depicted in FIG.\ref{fig:suppfig12}). This will be elaborated on later.

\subsection{Hilbert space structure and fragmentation}

Here, we present the connectivity structure of the Hilbert space of our system in sectors with a various number of particles. This reveals the geometric symmetry of our model that is not apparent when drawn in terms of an array of coupled effective Su–Schrieffer–Heeger (SSH) chains.

\begin{figure}[h]
	\includegraphics[width=0.6\linewidth]{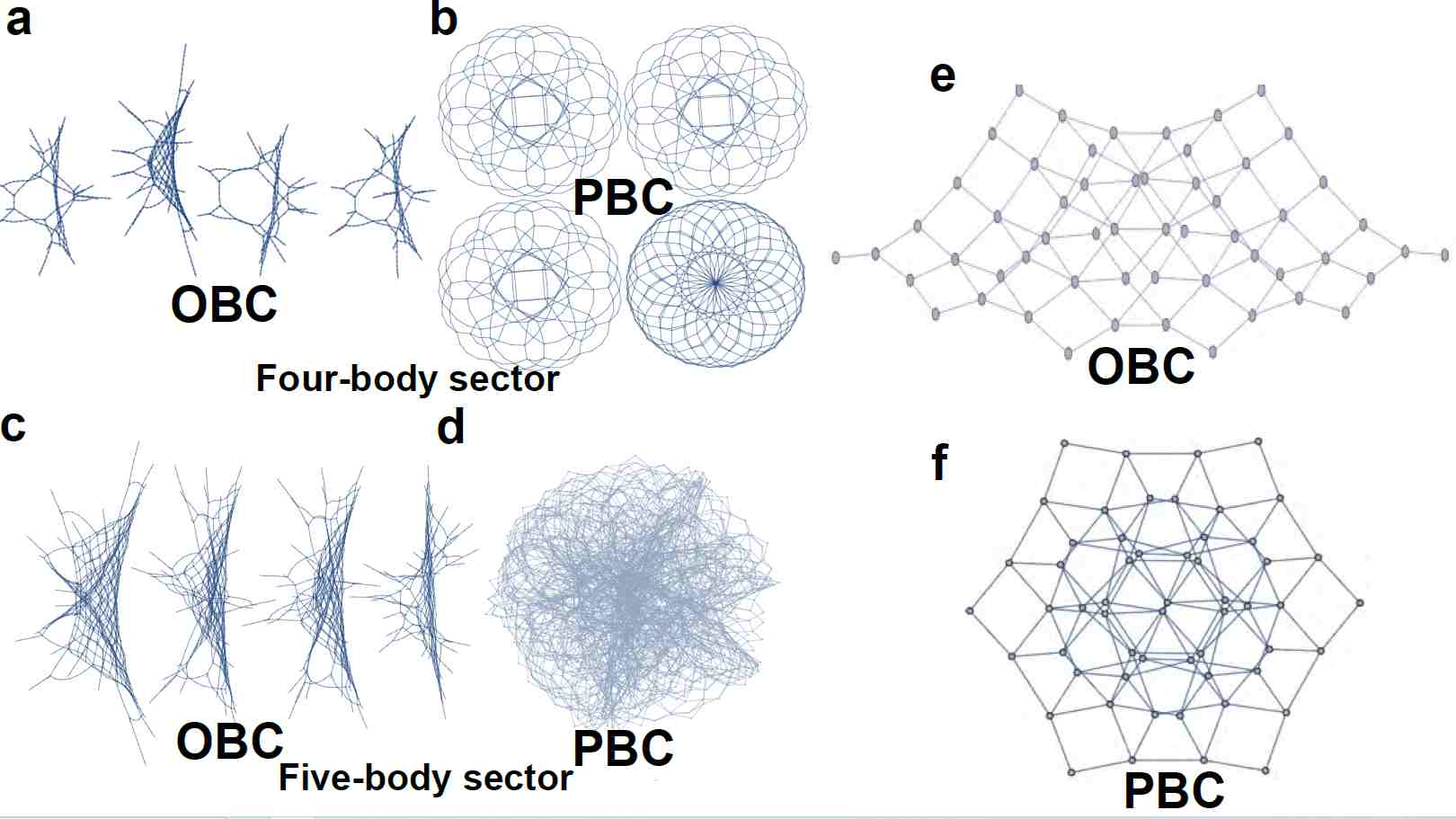}
	\caption{The Hilbert space connectivity graphs of our interacting Hamiltonian $H$ with $N$ bosons: \textbf{a} OBC, L=12, N=4, \textbf{b} PBC, L=12, N=4, \textbf{c} OBC, L=12, N=5, \textbf{d} PBC, L=12, N=5. For reference, also plotted are the graphs for the half-filled Bose-Hubbard model: \textbf{e} OBC, L=6, \textbf{f} PBC, L=6.}
	\label{fig:suppfig3}
\end{figure}

In our connectivity graphs, the tunneling between the two states, irrespective of actual amplitude or type, contributes to one effective hopping (See Fig.~\ref{fig:suppfig3} \textbf{a}-\textbf{b}).  The connectivity graphs reveal some degree of Hilbert space fragmentation induced by the interactions and particle statistics in our model for three or more particles (The Hilbert space graph consists of three or four disconnected subgraphs). In addition, it also reveals the nontrivial effects of boundary conditions: The Hilbert space structure of PBC vs. OBC graphs can be substantially different due to the non-local nature of some of the effective hoppings. %ei.e., Quantum many-body scar and many-body localization\cite{choi2019emergent,pietracaprina2019hilbert,zhao2020quantum,turner2018weak,turner2018quantum,turner2020correspondence}. The non-hermitian physics and scar state may give rise to the new dynamical regime, and create more novel states. 

As a comparison, we also plot the Hilbert space connectivity graph (Fig.~\ref{fig:suppfig3} \textbf{e}-\textbf{f})) of the half-filled Bose-Hubbard model with nearest hopping $t$, on-site interaction $U$, and potential $\mu$: $H_\text{Bose-Hubbard}=-t \sum_{\langle i, j\rangle} \hat{c}_{i}^{\dagger} \hat{c}_{j}+\frac{U}{2} \sum_{i} \hat{n}_{i}\left(\hat{n}_{i}-1\right)-\mu \sum_{i} \hat{n}_{i}$.

\subsection{Analytic ansatz for Interface clusters}

\begin{figure}[h]
	\centering
	\includegraphics[width=.6\linewidth]{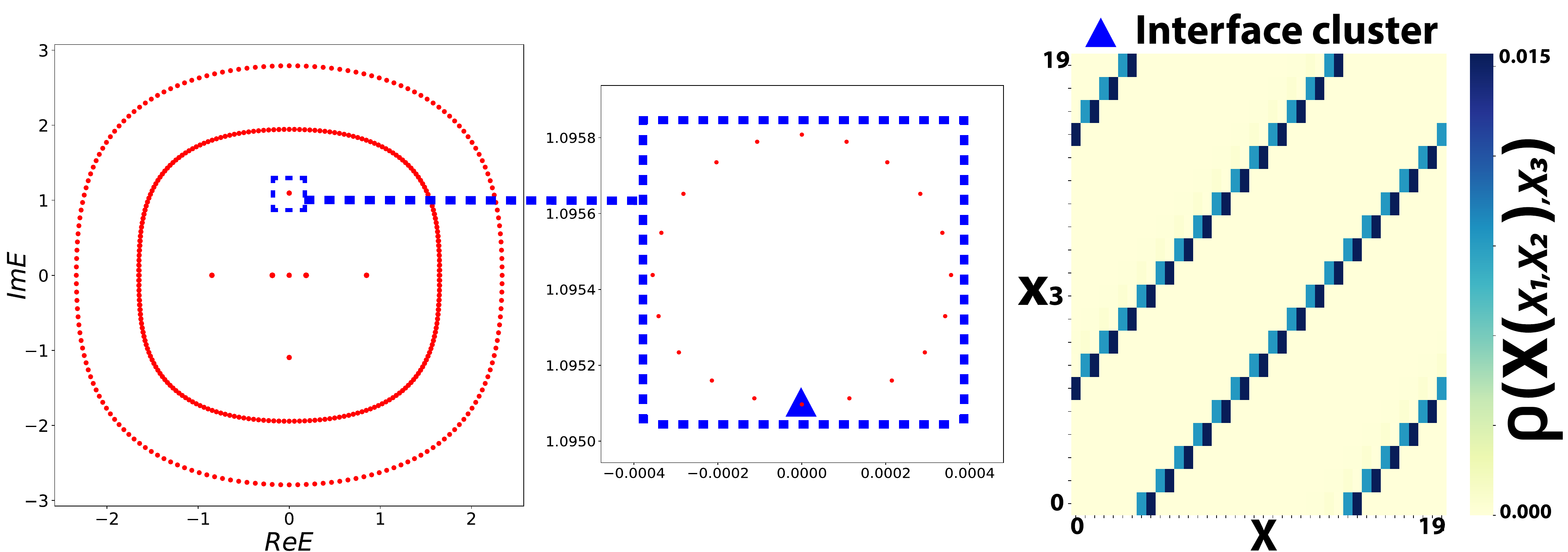}
	\caption{Representative 3-boson PBC interface cluster states with $t_{1}=1.0,t_{2}=0.2,\gamma=0.8,\,L=20$, which are almost degenerate in spectrum, and all localized along the interfaces in the effective lattice. Their average eigenenergy and spatial profile agrees excellently with the compact state ansatz.
	}
	\label{fig:suppfig4}
\end{figure}

Here we elaborate on the structure of special isolated interface clusters under PBCs (blue triangular dots in Fig.~\ref{fig:suppfig4}). We shall approximate them with the compact state ansatz $|i,j,k\rangle=c^{\dagger}_{i}c^{\dagger}_{j}c^{\dagger}_{k}|0\rangle$, which exist under PBCs:
\begin{equation}
	\left| \Psi\right\rangle _{com}=\sum_{i}^{L}(\psi_{-1}|i-1,i-1,i+3\rangle+\psi_{0}|i-3,i,i+3\rangle+\psi_{1}|i-3,i+2,i+2\rangle).
\end{equation}
Substituting it into our Hamiltonian $H$ (Eq.~\eqref{H}), we have
\begin{equation}
	\begin{aligned}
		(t_{2}-\gamma)(\psi_{1}+\psi_{-1})=E\psi_{0}\\
		(t_{2}+\gamma)\psi_{0}=E\psi_{1}\\
		(t_{2}+\gamma)\psi_{0}=E\psi_{-1}.
	\end{aligned}
\end{equation}
%From the terms $\psi_{-1}|(i-7,i-7),i-3>$ and $\psi_{1}|i-3,(i+2,i+2)>$, for PBC model, we can assume 
%\begin{equation}
%\begin{aligned}
%\psi_{1}^{2}=\kappa \psi_{-1}^{2}
%\end{aligned}
%\end{equation}
We get the solution
\begin{equation}
	\begin{aligned}
		\psi_{1}^{2}=\psi_{-1}^{2},\quad E=\pm\sqrt{2(t_{2}^{2}-\gamma^{2})},
	\end{aligned}
	\label{E}
\end{equation}
which turns out to be an excellent approximation of our interface cluster states. In the case shown in Fig.~\ref{fig:suppfig4}, $E_{numerical}\approx\pm1.0955491i$ for the numerical interface cluster, and $E_{analytical}=\pm\sqrt{2(t_{2}^{2}-\gamma^{2})}\approx\pm1.0954451i$ for the compact state ansatz $\left| \Psi\right\rangle _{com}\approx \left| \Psi\right\rangle _{interface}$. 
\clearpage
\subsection{Comparison between various types of cluster states}
In Fig.~\ref{fig:suppfig5} below, we provide more detailed comparisons between representative states of distinct types, plotted in both the physical and effective lattices.

\begin{figure}[h]
	\centering
	\includegraphics[width=.95\linewidth]{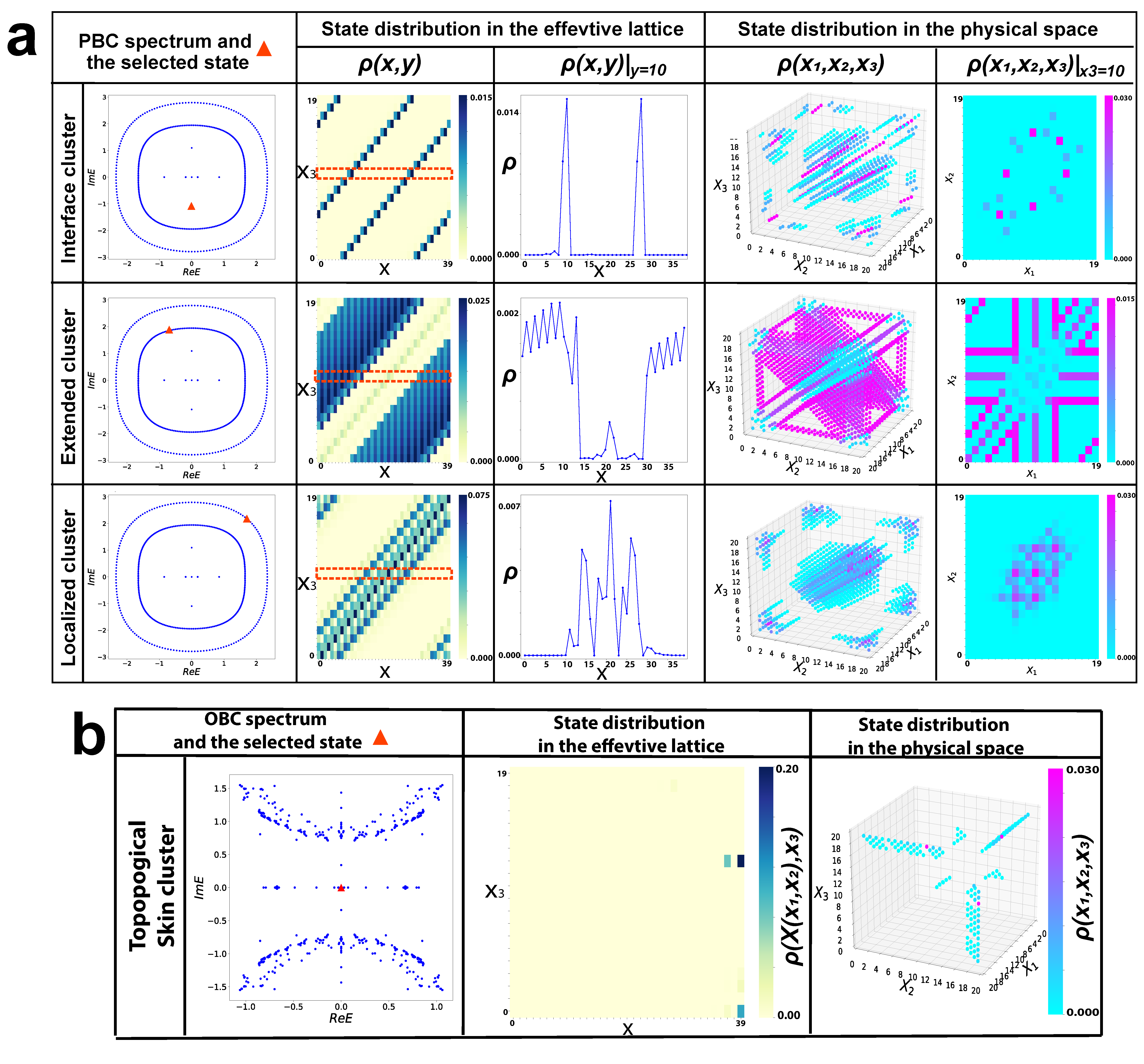}
	\caption{3-boson skin cluster states of $H$ with parameters $t_{1}=1.0,t_{2}=0.2,\gamma=0.8,L=20$. \textbf{a} PBC case, from left to right: The PBC spectrum; the state distribution in the effective lattice $\rho(X,x_{3})$ of the selected state (orange triangle), and its density plot $\rho(X,x_{3})|_{x_{3}=10}$ along a cross-section; its spatial distribution  $\rho(x_{1},x_{2},x_{3})$ in physical space (the sites with $\rho(x_{1},x_{2},x_{3})<0.0001$ are not plotted for clarity), and the same data along the cross-section $\rho(x_{1},x_{2},x_{3})|_{x_{3}=10}$. Indeed, localized clusters are the most tightly clustered; interface clusters comprises configurations at the literal interface of localized clusters; and extended clusters frequently involve one arbitrarily far boson. 
		\textbf{b} OBC case, from left to right: the OBC spectrum; the state distribution in the effective lattice $\rho(X,x_{3})$ of the selected zero mode state (orange triangle); its spatial distribution $\rho(x_{1},x_{2},x_{3})$ in physical space, revealing essentially no dependence on one of the particle coordinates i.e. it is a two-body phenomenon.
	}
	\label{fig:suppfig5}
\end{figure}

\clearpage
\subsection{PBC spectral loop structure for skin cluster states}
\begin{figure}[h]
	\centering
	\includegraphics[width=0.99\linewidth]{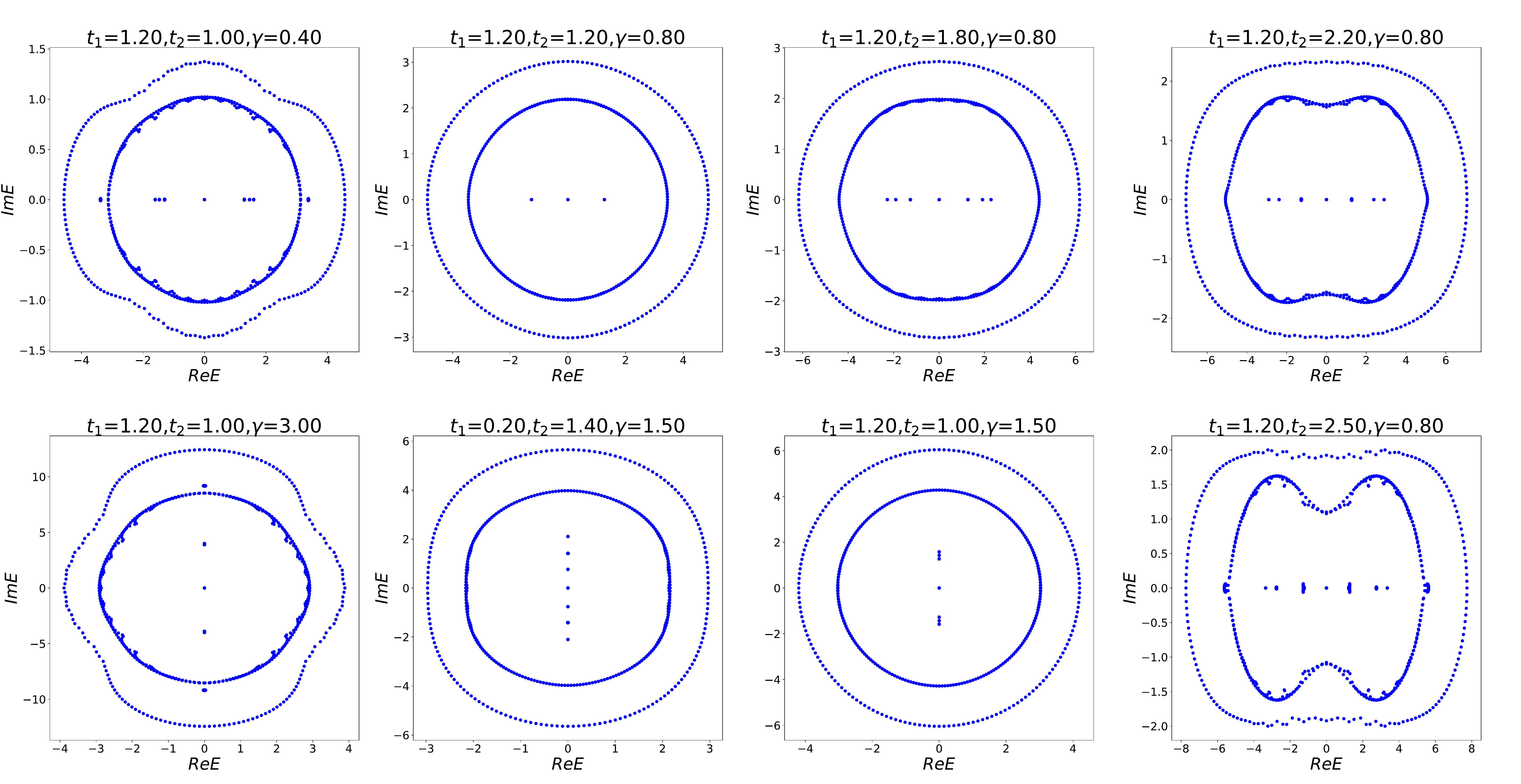}
	\caption{3-boson PBC spectra with different $t_{1},t_{2},\gamma$. The two concentric loops always remain, even though their detailed shape changes. The length of the physical chain is $L=20$.}
	\label{fig:suppfig6}
\end{figure}

Under a wide range of  $t_{1},t_{2},\gamma$ parameters, the PBC spectrum in the three-body sector exhibits distinct concentric loops with states on the outer and inner loops exhibiting utterly different behavior. The exact shapes of these loops depend on their parameters (see Fig.~\ref{fig:suppfig6}), but their clear separation i.e. loop gap is rather robust. Previously, we have already classified these states into localized clusters (outer loop), extended clusters (inner loop), and interface clusters (isolated state hosted by Eq. \eqref{E}).

\begin{figure}[h]
	\centering
	\includegraphics[width=0.85\linewidth]{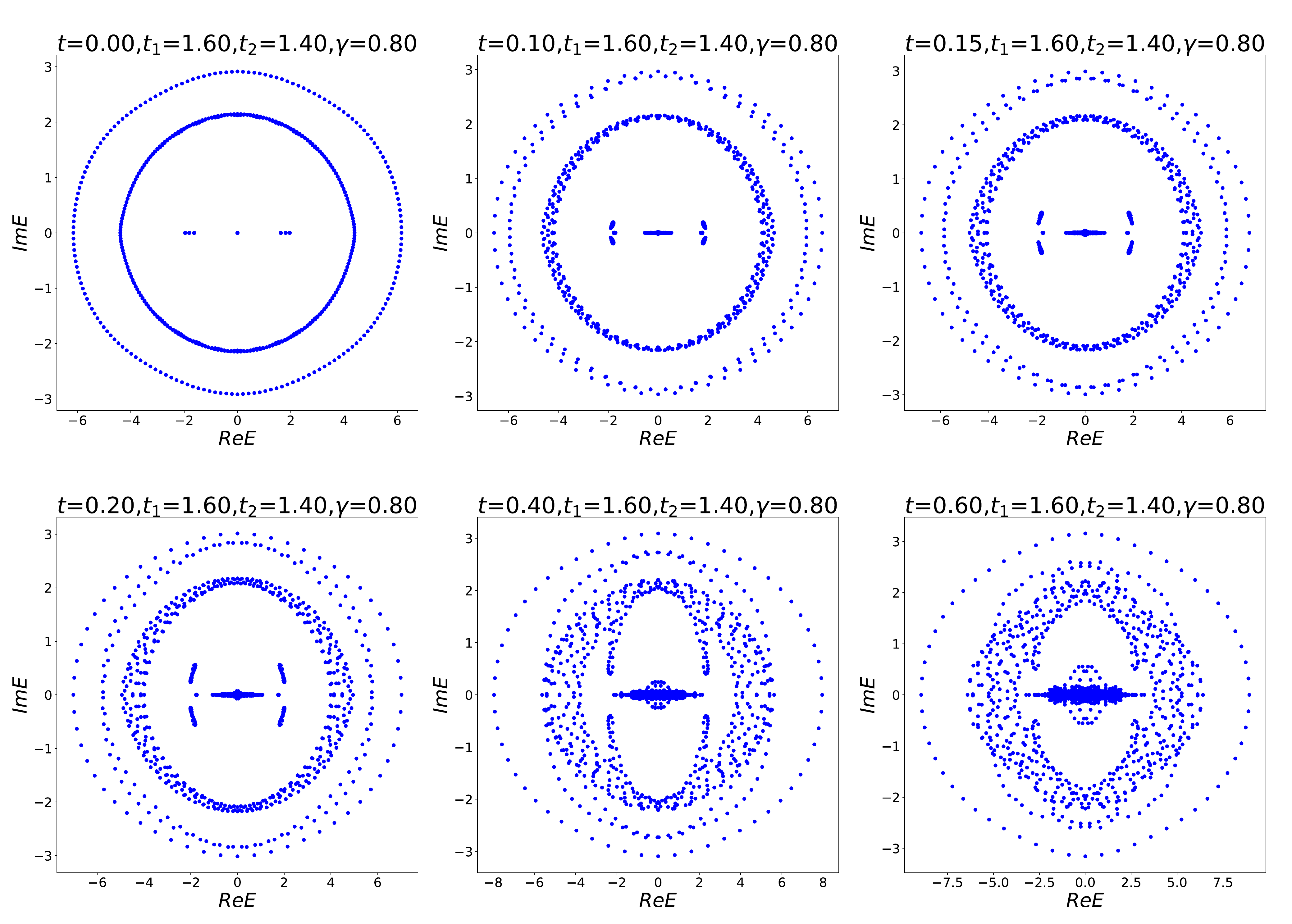}
	\caption{The three-body PBC spectrum of $H'$ with fixed $t_{1},t_{2},\gamma$ and different values of single-body dispersion $t$. The length of the physical chain is $L=20$. Even though the inner loop (extended clusters) and isolated inner states (interface clusters) start to merge at sufficiently large $t$, the outer loop (localized clusters) remains untouched, separated from the other (inner) states by a robust loop gap.}
	\label{fig:suppfig7}
\end{figure}

\begin{figure}[h]
	\centering
	\includegraphics[width=0.9\linewidth]{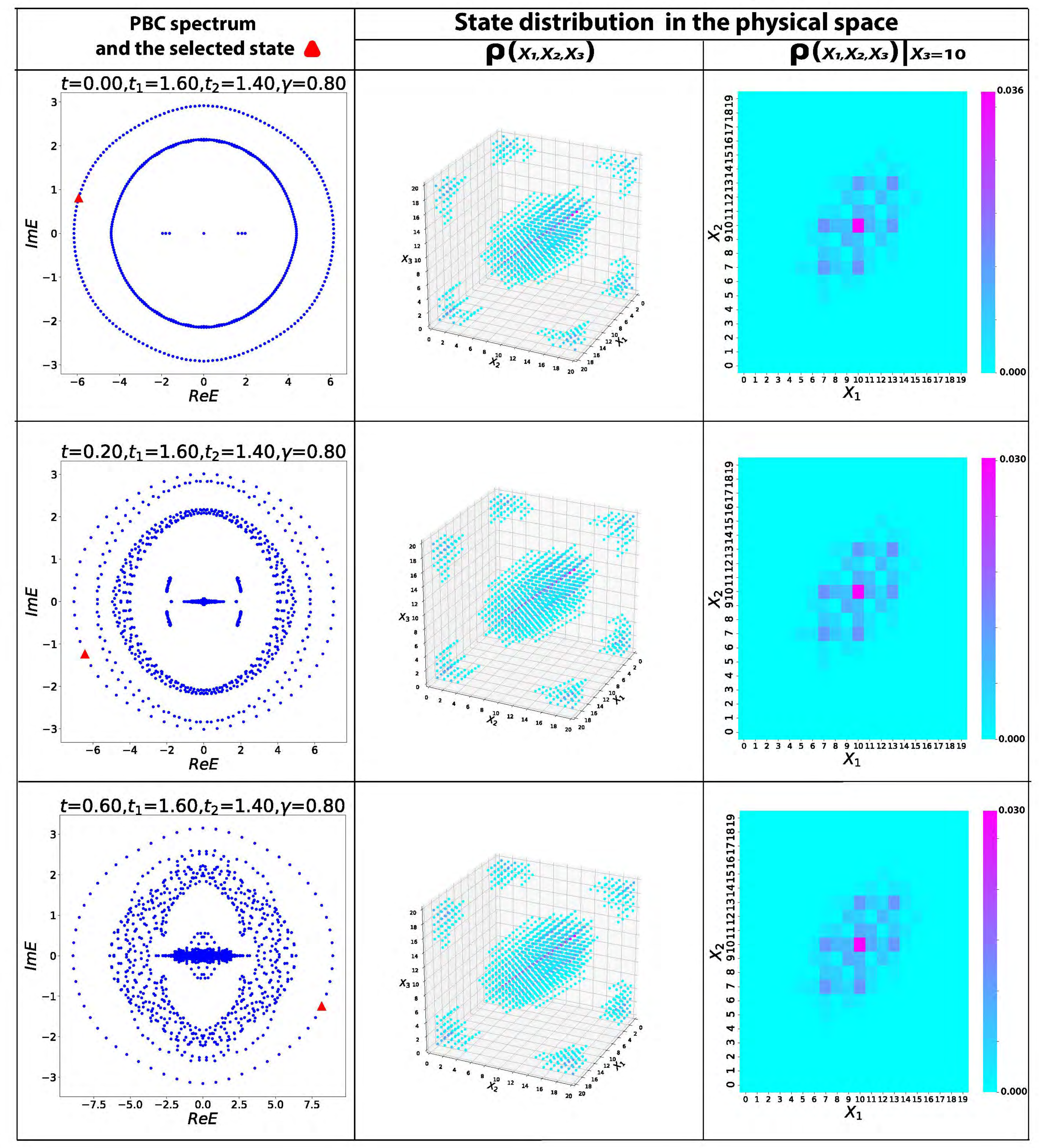}
	\caption{Robustness of the localized clusters on the outer spectral loop, subject to different $t$. Left: The three-body PBC spectrum with $t=0,0.2,0.6;t_{1}=1.6,t_{2}=1.4,\gamma=0.8$. The length of the physical chain is $L=20$. Middle: The physical spatialdistribution of the selected state (red triangle), and Right: its corresponding cross-section at fixed $X_{3}=10$. In the 3D state distribution plot, the sites with $\rho(x_{1},x_{2},x_{3})<0.0001$ are left out for clarity. Clearly, the localized clusters remain almost invariant across different $t$.}
	\label{fig:suppfig8}
\end{figure}

To further probe the robustness of the loop gap, we add physical nearest neighbor Hermitian hoppings to our original interacting Hamiltonian $H$:
\begin{equation}
	H^{\prime}=H+ t\sum_{i}(c^{\dagger}_{i+1}c_{i}+h.c.)
\end{equation}
In other words, we perturb our original model $H$ with single-body dynamics with strength $t$. Fig.~\ref{fig:suppfig7} presents the PBC spectrum of three boson $H$' with increasing $t$. We find that when increasing $t$, some states move towards the zero energy origin, blurring the inner extended clusters loop. At sufficiently large $t$, the isolated interface clusters acquire nontrivial dynamics and start to merge with the extended clusters. However, the localized clusters (outer loop) remain stable even when $t$ is comparable to the original interaction terms in $H$. This can be further verified in the spatial distribution of the localized clusters upon increasing $t$ (see Fig.~\ref{fig:suppfig8}), which indeed remain almost unchanged. %, from the comparison between the cases $t=0$ and $t\neq0$, we confirm that some states on the outer band exhibit robustness that there are still some states on the outer band. Morevore, the $t$ does not affect the behavior of the outer-band state distribution in the physical space(See the FIG.~\ref{suppfig8}).

%\newpage

\clearpage
\subsubsection{Correlation density functions and the interaction-induced loop gap}
In our two-body correlated hopping model $H$, the clustering properties of the bosonic eigenstates constitute the most important results. The extent of pair clustering of a given state $\psi_m$ can be measured by the density-density correlation function 
\begin{equation}
	\begin{aligned}
		C_{m}(x_{i},x_{j})=\left|\left\langle \psi_{m} \left| c^{\dagger}_{i}c_{i}c^{\dagger}_{j}c_{j} \right|\psi_{m} \right\rangle  \right|.
	\end{aligned}
\end{equation}
Since $H$ obviously requires two bosons to occupy the same site either before or after they hop, we trivially have strong peaks in $C_m(x_i,x_j)$ for $i=j$. To better illustrate what happens when $x_i\neq x_j$, we shall omit plotting $C_m(x_i,x_i)$ in the following plots.

\begin{figure}[h]
	\centering
	\includegraphics[width=\linewidth]{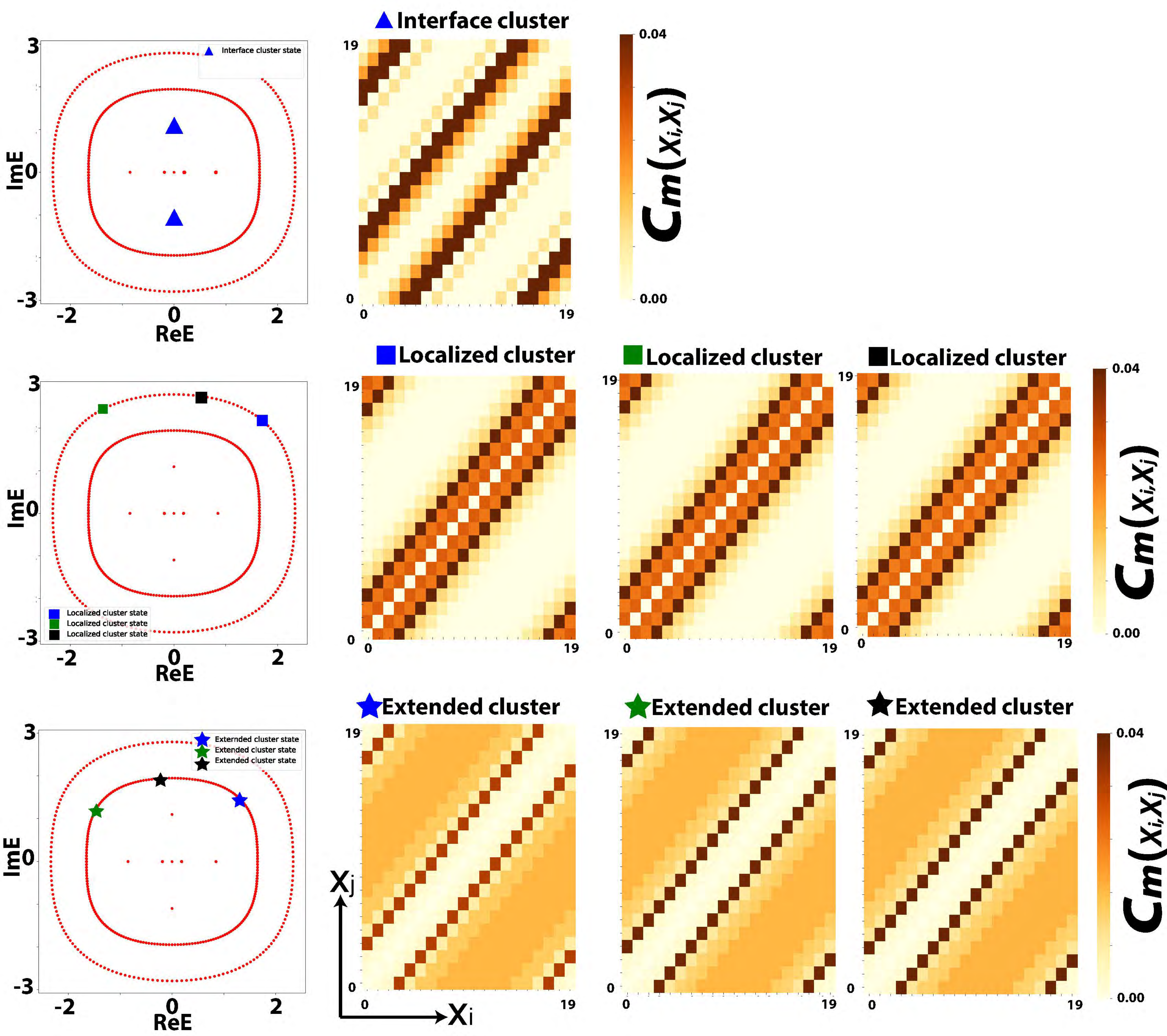}
	\caption{The density-density correlation function $C_{m}(x_{i},x_{j})$ for illustrative PBC 3-boson clusters of various types, with $t_{1}=1.0,t_{2}=0.2,\gamma=0.8,L=20$. There is little variation between the two-body clustering behaviors of different states of the same type, even though their eigenenergies can be completely different (but still lying within the same loop).}
	\label{fig:suppfig9}
\end{figure}

\begin{figure}[h]
	\includegraphics[width=.8\linewidth]{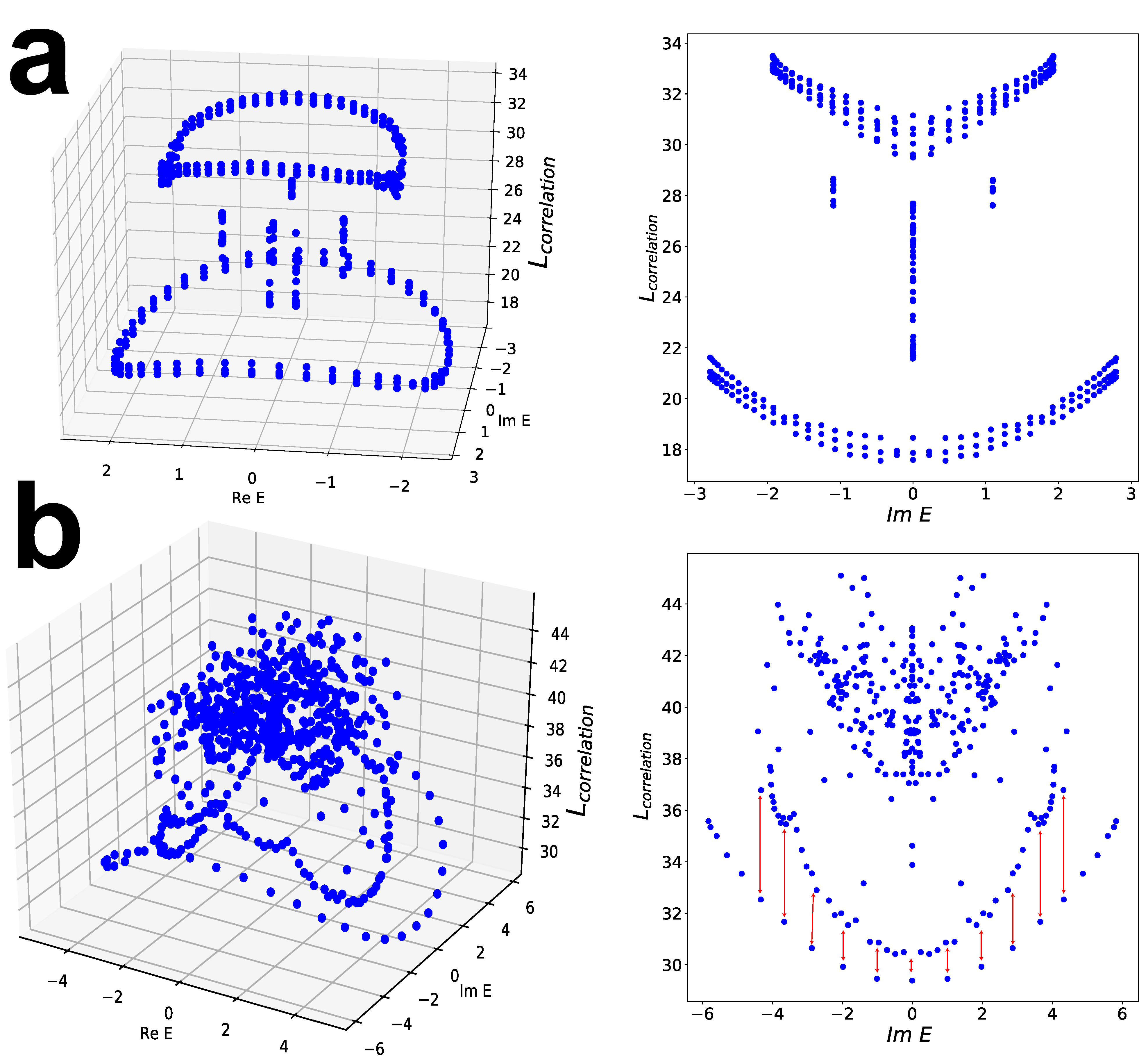}
	\caption{Plots of the pair correlation lengths $L_\text{correlation}$ of all eigenstates as a function of the complex energy (Left), and as a function of the imaginary energy i.e. with collapsed real energy (Right). Shown are \textbf{a} 3-boson clusters and \textbf{b} 4-boson clusters, both corresponding to parameters $t_{1}=1,t_{2}=0.2,\gamma=0.8$, but with $L=15$ and $L=10$ respectively. The contrast in $L_\text{correlation}$ is clear for the 3-boson clusters; for 4-boson clusters, it is also quite evident when viewed along the imaginary energy axis. The outer loop splits into two loops with slightly different pair correlation lengths, as indicated by the red vertical lines.%	The plot of the localization length in the $ReE-ImE-L_{corrlation}$(right), and its section in $ImE-L_{correlation}$(left) for:(a)three-body model with $t_{1}=1,t_{2}=0.2,\gamma=0.8,L=15$; (b)four-body model with $t_{1}=1,t_{2}=0.2,\gamma=0.8,L=10$.
	}
	\label{fig:suppfig10}
\end{figure}

\clearpage
From Fig.~\ref{fig:suppfig9} above, what differentiates the outer (localized cluster states) and the inner loop states (extended cluster states) is their clustering properties. Qualitatively, the localized states are contributed more by short-range correlations, while the extended states are contributed more by the long-range correlations. To better quantify that, we introduce the pair correlation length of a given state $\psi_k$:
\begin{equation}
	\begin{aligned}
		L_\text{correlation}=\sum_{i, j}\left|\langle\psi_{k}\right|\left(\hat{x}_{i}-\hat{x}_{j}\right) \hat{n}_{i} \hat{n}_{j}\left|\psi_{k}\rangle\right|
	\end{aligned}
\end{equation}
where the $\hat{x}_{i}$ is the position operator and the $\hat{n}_{i}=c^\dagger_ic_i$ is the occupation number operator for the $i$-th site. Fig.~\ref{fig:suppfig10} shows that the pair correlation length is clearly longer for the inner spectral loop (extended clusters), both for 3-boson and 4-boson clusters. The magnitude of $L_\text{correlation}$ scales like $\langle \hat n_i \hat n_j\Delta x \rangle$, and is largest if the state contains two dense ``clumps'' that are widely separated.

%\noindent{\textit{Section on loop gap... --}} \red{CH: please explain the loop gap in terms of the effective quasiparticles}
%
%The phenomenon of the boson cluster in our model gives rise to the concept of the composite particle. We can treat this composite particle as the special excitation. This formalism suggests that the excitation can induce the energy gap, and the higher-energy state owns higher density of composite-particle excitation. To confirm our assumption, we express the density as

\begin{figure}[h]
	\includegraphics[width=.9\linewidth]{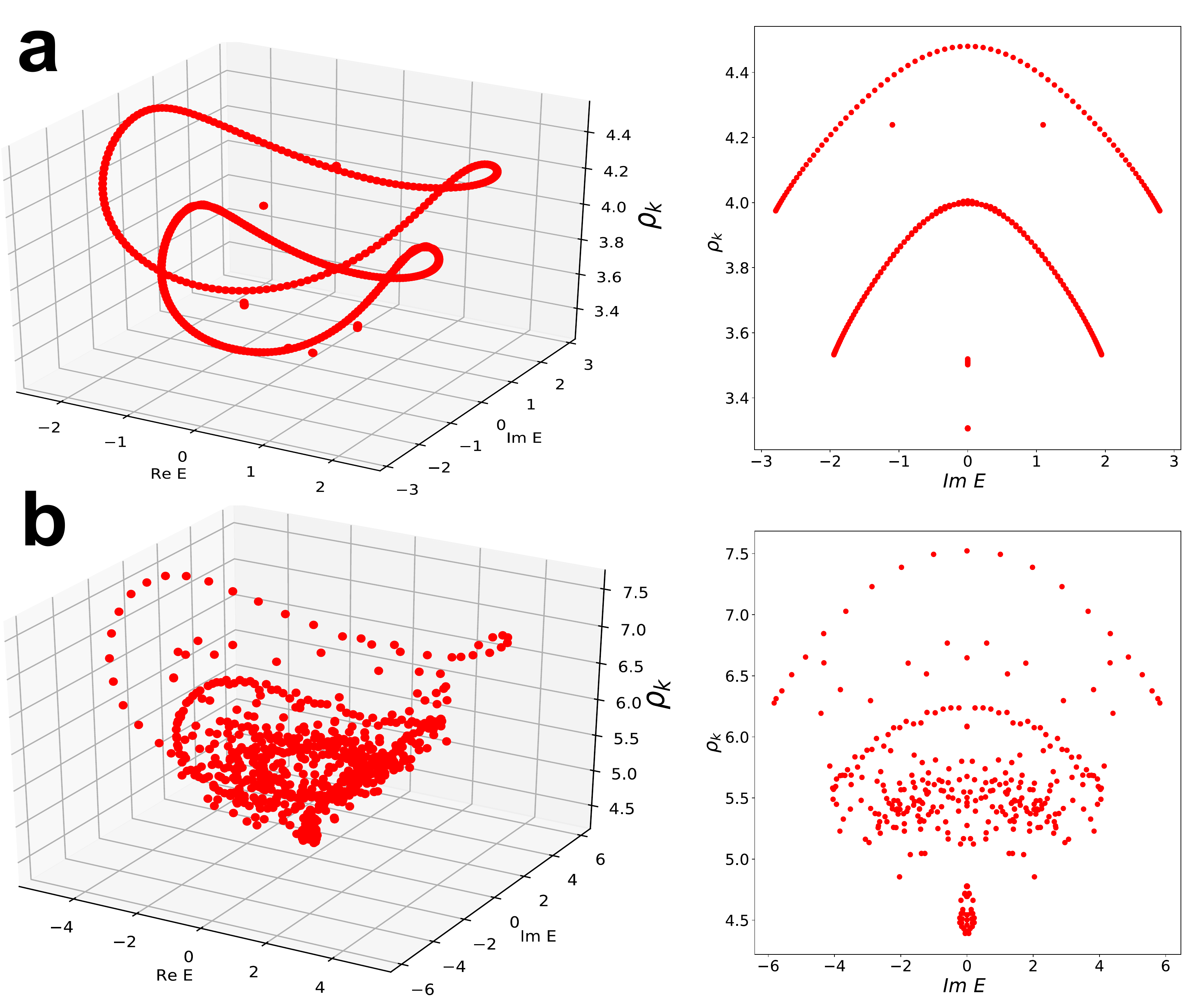}
	\caption{The cluster density $\rho_k$ plotted in the complex energy plane (Left) and  also the imaginary energy (Right), so as to resolve the different loops more distinctively. Plotted in \textbf{a} and \textbf{b} are the 3-boson and 4-boson cases with $L=20$ and $L=10$ respectively, both with parameters $t_{1}=1.0,t_{2}=0.2,\gamma=0.8$.  }
	\label{fig:suppfig11}
\end{figure}

\clearpage
As a complementary measure of clustering, we also introduce the cluster density $\rho_k$ for a given state $\psi_k$:
\begin{equation}
	\begin{aligned}
		\rho_{k}=\sum_{i}\left\langle \psi_{k} \left| n^{2}_{i}\right|\psi_{k} \right\rangle.
	\end{aligned}
\end{equation}
It measures the density of clusters for a particular state, and will be maximal for states where more bosons overlap at the same site. Fig.~\ref{fig:suppfig11} presents the distribution of the cluster density for three-boson and four-boson PBC clusters. There is a clear gap from the outer to inner loops, indicating significantly more "clusters" on the outer loop cluster states. Intermediate states emerge between the outer and inner spectral loops in the four-boson case, delineating a saddle surface. Overall, the cluster density is evidently larger for outer loop states with large energies, consistent with the discussion in the main text.

\newpage

\section{Supplementary Note 2: Clustering behavior in 3-boson, 4-boson, and 5-boson clusters}\label{Note2}
\subsection{Effective interface in high dimensional Hilbert spaces belonging to many-boson sectors}
\begin{figure}[h]
	\centering
	\includegraphics[width=.73\linewidth]{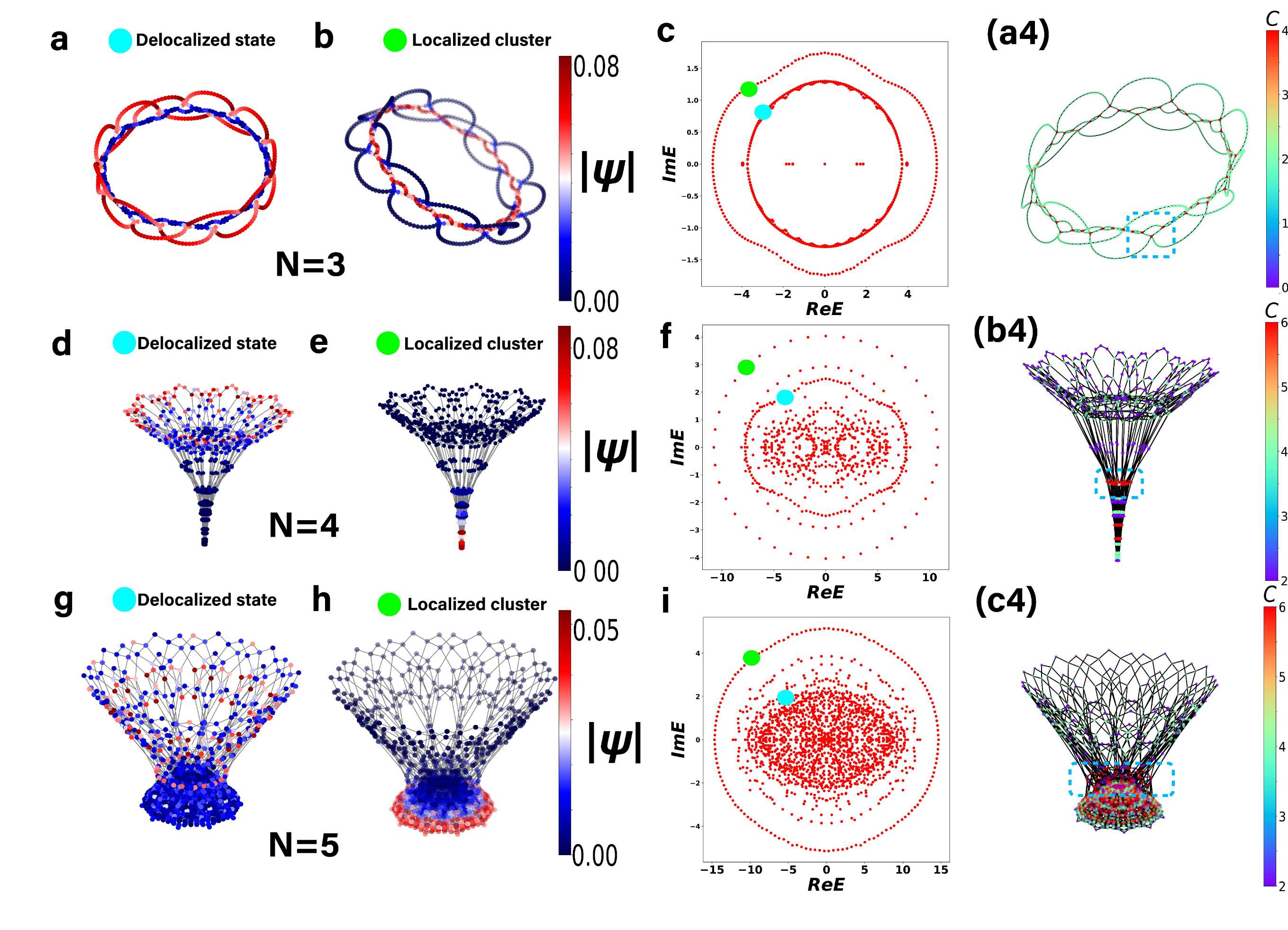}
	\caption{Detailed illustration of the effective interface and Hilbert space graph states given in the main text for the selected states in N-boson sector unde PBCs: \textbf{a}, \textbf{d} and \textbf{g}: delocalized states (blue dots in \textbf{c}, \textbf{f} and \textbf{i}); \textbf{b}, \textbf{e} and \textbf{h}: localized cluster modes (green dots in \textbf{c}, \textbf{f} and \textbf{i}). The colorbar in blue-red scheme represents density of the state in the graph. Parameters are $t_{1}=1.4,t_{2}=1.2,\gamma=0.5$. The size and filling are \textbf{a}, \textbf{b} L=20, N=3; \textbf{d}, \textbf{e} L=10, N=4; \textbf{g}, \textbf{h} L=10, N=5. The localized skin cluster is found on the outer spectral loop and the delocalized state lies within the loop gap, for all the cases. This generic behavior arises from the effect of effective interface (see details in FIG. \ref{fig:suppfig13}).}
	\label{fig:suppfig12}
\end{figure}

In this section, we discuss how the effective interface in the Hilbert space for 3-boson can be naturally embedded in the Hilbert space when there are $N>3$ bosons. First, we focus on the localized cluster mode. For the three-boson sector, the localized cluster mode under PBCs appears in the states at the outer spectra loop (see the states on the outer loop in FIG.\ref{fig:suppfig11}). 
For our N-boson sector under PBCs in FIG.\ref{fig:suppfig12}, we find that the spectra exhibit a similar structure, and the states that are localized clusters (green dots) in presence of non-Hermiticiy appear in the outer spectral loops. Here, we demonstrate with $N=3,4,5$ particles, which indicates that the localized cluster mode is not a special case in the 3-boson sector. This is evidence that the filling fraction of the boson does not affect the skin cluster localization behavior under PBCs. In our analysis of the 3-boson model, we find that different types of states under PBCs and OBCs are the result of an effective interface. In FIG.\ref{fig:suppfig12}, we show that the skin cluster localization as localized cluster mode exist under $N=3,4,5$. Therefore, apart from the given localized cluster mode, other types of modes should also exist when we have the other bosonic fillings.

\begin{figure}[th!]
	\centering
	\includegraphics[width=0.6\linewidth]{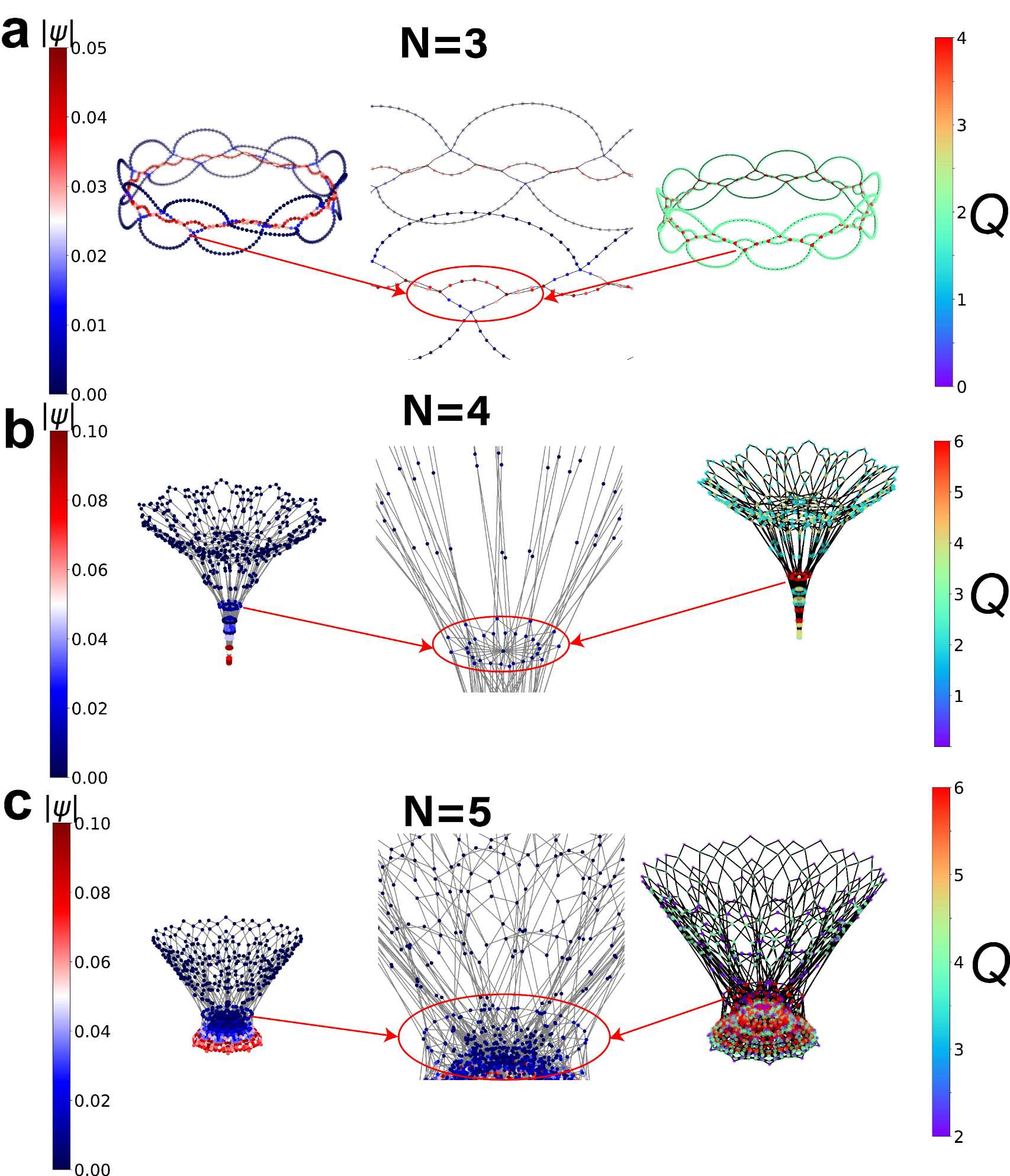}
	\caption{Effective interfaces in \textbf{a} 4-boson sector and \textbf{c} 5-boson sector in FIG.\ref{fig:suppfig12} under PBCs. The effective interfaces separate the states with relatively high and low connectivity. The connectivity $Q$ is the number of neighbors at each node, and $|\psi|$ represents the density of the state in the graph. The effective interfaces are shown in the zoomed in plot which separate the graph into regions with different connectivity. The remarkable feature is that the effective interface appears near the nodes red, which has maximal number of neighbors.}
	\label{fig:suppfig13}
\end{figure}
Inspired by the above localization behavior on the Hilbert space graph, in FIG.\ref{fig:suppfig13}, we give details on how an effective interface, using one particular graph as an example. To differentiate nodes in the graph, we label the nodes with different colors in terms of their number of nearest neighbors $Q$ (graph connectivity). According to the graph in FIG.\ref{fig:suppfig13}, there exist nodes with maximal number of neighbors that separate the graph into regions with different connectivity. Correspondingly, on the LHS, we check the state density in the graph selected from the localized states in FIG.\ref{fig:suppfig12} and find the localization exactly appears in the region that is separated by the nodes with maximum neighbors. Thus, these nodes play the role of the effective interface in the graph.
\newpage
\begin{figure}[th!]
	\centering
	\includegraphics[width=0.5\linewidth]{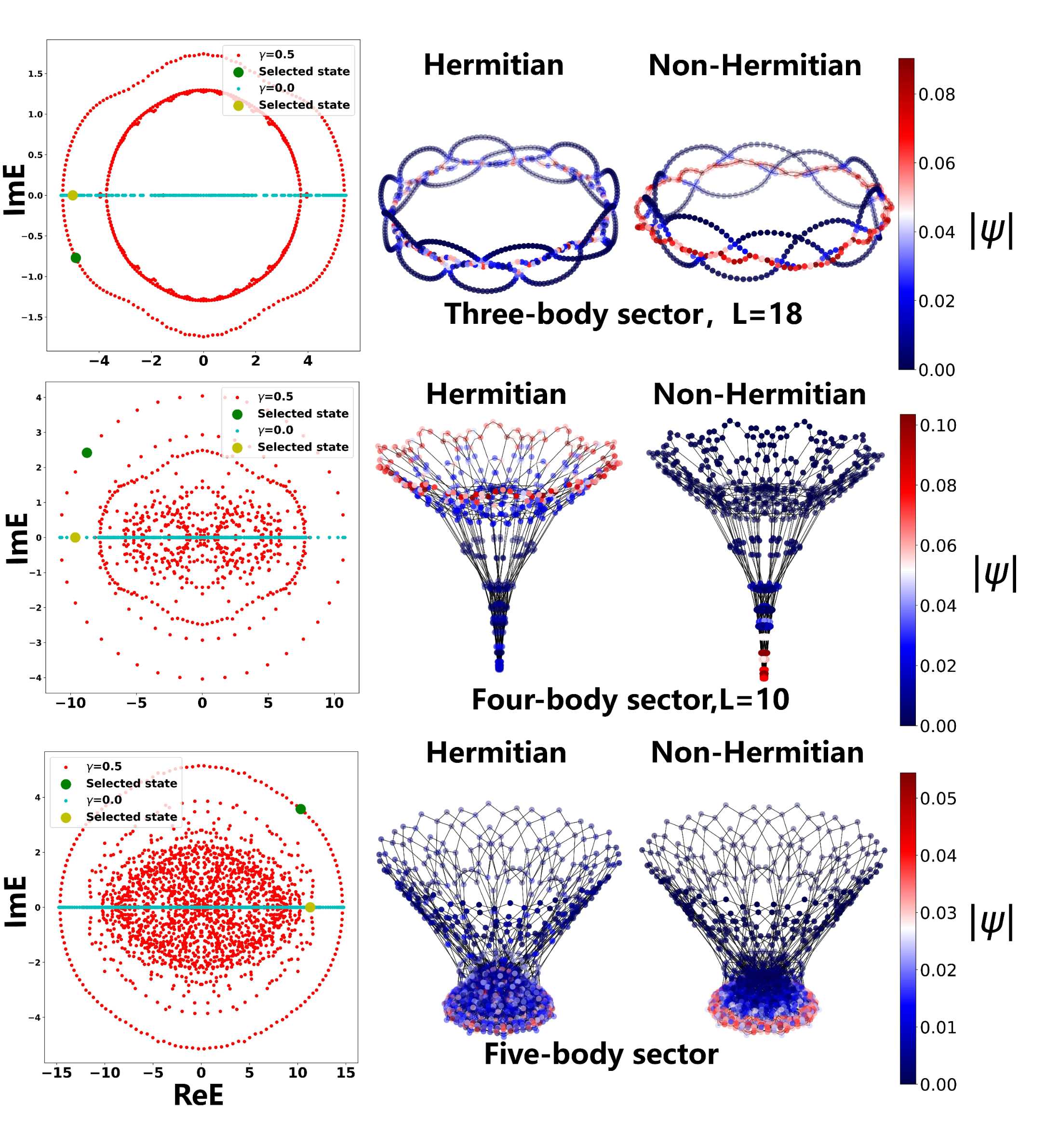}	
	\caption{Selected PBC eigenstates in the Hilbert space graph, contrasting the behavior of Hermitian and non-Hermitian cases. The colorbar represents the density of the state in the graph. The state in the green square is the state for the non-Hermitian case, and the state in the yellow square is for the Hermitian case. Parameters are $t_{1}=1.4,t_{2}=1.2$, $\gamma=0.0$ for Hermitian case (blue dots in spectra) and $\gamma=0.5$ for non-Hermitian case (red dots in spectra). There exists the skin localization in the region that is separated by the effective interface in FIG.\ref{fig:suppfig13} when $\gamma\neq 0$. In the three-boson sector, we choose the extended skin cluster state. Similarly, for four and five-boson sectors, we also select the state on the outer spectral loop, and these selected state also exhibits the skin localization in the graph which indicates the existence of extended skin cluster. The existence of skin clusters crucially require non-Hermiticity.}
	\label{fig:suppfig14}
\end{figure}
In FIG.\ref{fig:suppfig14}, we find that in these restricted regions with higher connectivity shown in FIG.\ref{fig:suppfig13}, and our model only exhibits the skin localization in presence of non-Hermiticity, implying that that skin clustering requires non-Hermiticity as well, and is not the simple effect of the geometry of the graph.
\newpage
\begin{figure}[th!]
	\centering
	\includegraphics[width=0.5\linewidth]{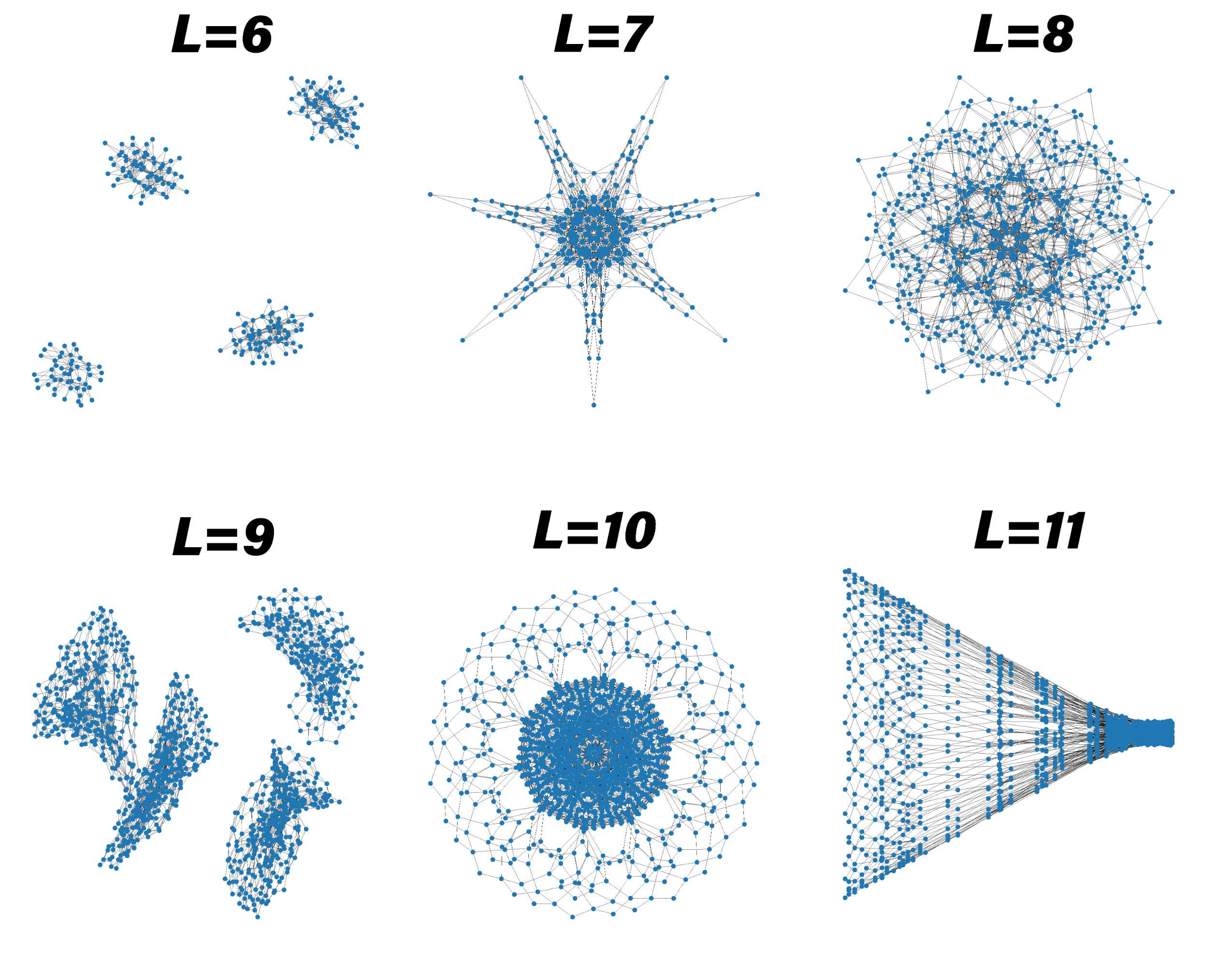}
	\caption{Hilbert space graph of 5-boson cases under PBCs, for different system size $L$. Hilbert space fragmentation appears when $L=6,9,12$ i.e. multiples of 3. For the 5-boson sector with L=10 in FIG.\ref{fig:suppfig13}(c), which is viewed from above, the effective interface can be clearly seen to separate the dense and sparse graph regions.}
	\label{suppfig15}
\end{figure}
To explain the fragmentation in FIG.\ref{suppfig15}, we consider one minimal N-boson model with L=3, and express states in Hilbert space as $\left| n_{1},n_{2},n_{3}\right\rangle$ ($\sum_{i}n_{i}=N$ and $N>3$
) where $n_{i}$ is occupation. We find that $\left| N,0,0\right\rangle$, $\left| 0,N,0\right\rangle$, $\left| 0,0,N\right\rangle$, and $\left| N-2,1,1\right\rangle$ are in four disconnected sectors under our Hamiltonian. Thus, if we have L=3n under PBC, the above four disconnected sectors just get expanded without additional sectors. As a result, in FIG.\ref{suppfig15}, when we have L=6 and L=9, we can obtain four disconnected sectors. For the three-boson model, the state $\left| 1,1,1\right\rangle$ is one trivial state.
\subsection{Orientated Quantum walk in high dimensional Hilbert space}
In this section, to showcase that the effective interface gives rise to inevitable state clustering accumulation, we consider the quenching dynamics in graph on one randomly prepared inital state $\left|\psi(0)\right\rangle$
\begin{equation}
	\begin{aligned}
		\left|\psi(t)\right\rangle=\frac{e^{-it\hat{H}}\left|\psi(0)\right\rangle}{\left\| \left|\psi(t)\right\rangle\right\| },
	\end{aligned}
\end{equation}
where $\left|\psi(t)\right\rangle$ is normalized.
\begin{figure}[th!]
	\centering
	\includegraphics[width=0.5 \linewidth]{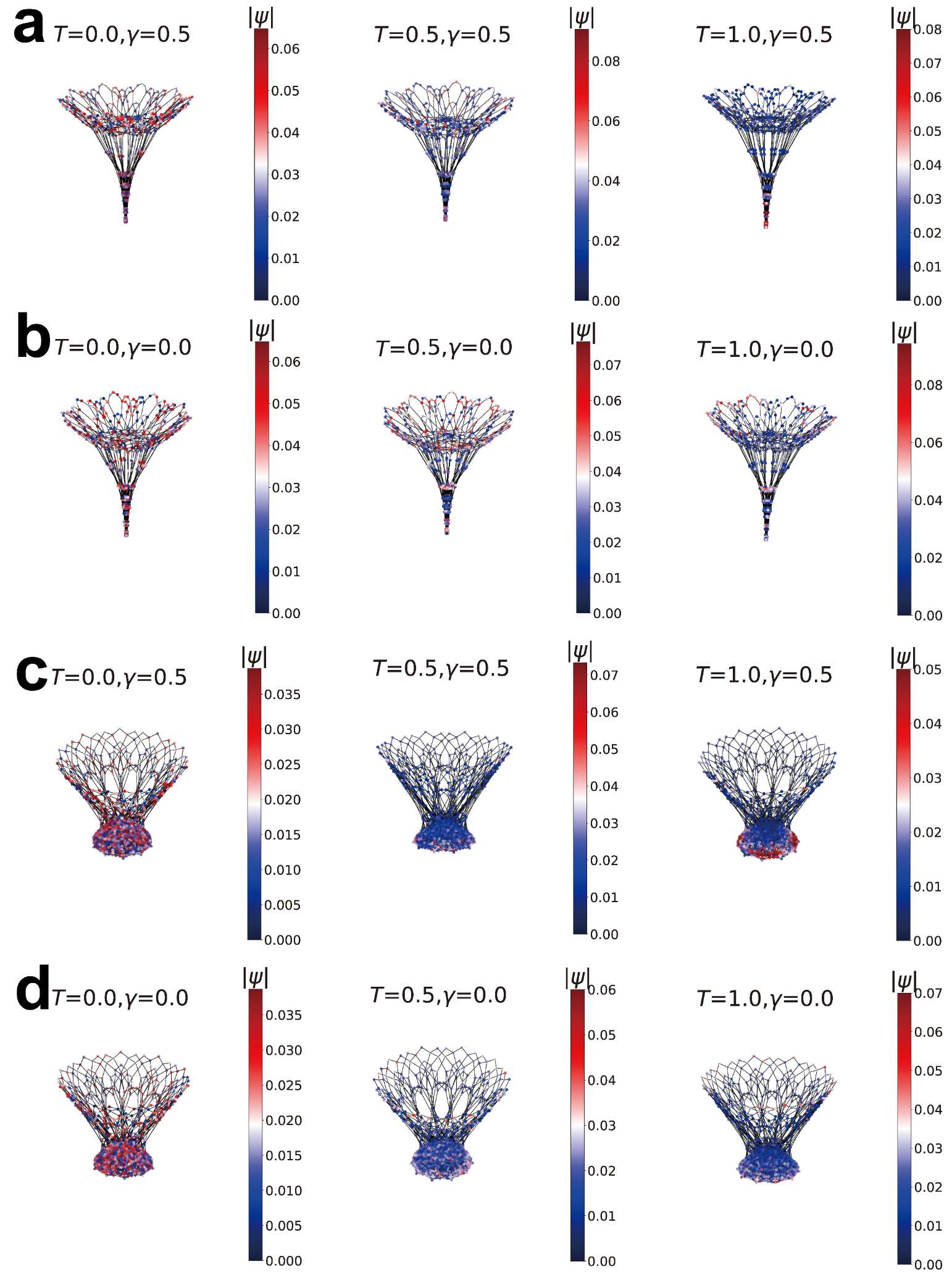}
	\caption{Dynamical evolution on a randomly prepared state in the Hilbert graph. \textbf{a} and \textbf{b} are for four-body sectors in a system with size $L=10$, and  \textbf{c} and \textbf{d} are for five-body sector in size $L=10$. The colorbar represents density of the state in the graph. For plots in \textbf{a}-\textbf{d}, the first plot is the initial state. We set $t_{1}=1.4, t_{2}=1.2$. We compare the evolution with ($\gamma=0$ for \textbf{b} and \textbf{d}) or without non-Hermiticity ($\gamma=0.5$ for \textbf{a} and \textbf{c}), and find the non-Hermiticity leads the skin localization shown in FIG.\ref{fig:suppfig13} during evolution.}
	\label{fig:suppfig16}
\end{figure}
\newpage
In FIG.\ref{fig:suppfig16}, we prepare the initial state that the state density randomly occupies the nodes in the graph, and compare quenching dynamics in two scenarios with or without non-Hermiticity. As indicated in FIG.\ref{fig:suppfig12}, there exist the eigenstates with skin localization in the region separated by an effective interface. Governed by non-Hermiticity, the state tends to evolve to the region orientated in FIG.\ref{fig:suppfig16} for both four and five-boson sectors during quenching dynamics. Due to the randomness of the initial state, these states can always meet the effective interface in the graph, and this effective interface can play a role as the boundary. Thus, we demonstrate that according to the behavior of dynamics in FIG.\ref{fig:suppfig16}, non-Hermitian skin effect can also exist in the graph with the effective interface. Naturally, for the sector with more particles, the effective interface in FIG.\ref{fig:suppfig12} is general and the corresponding skin localization is expected.
\begin{figure}[h!]
	\centering
	\includegraphics[width=0.9\linewidth]{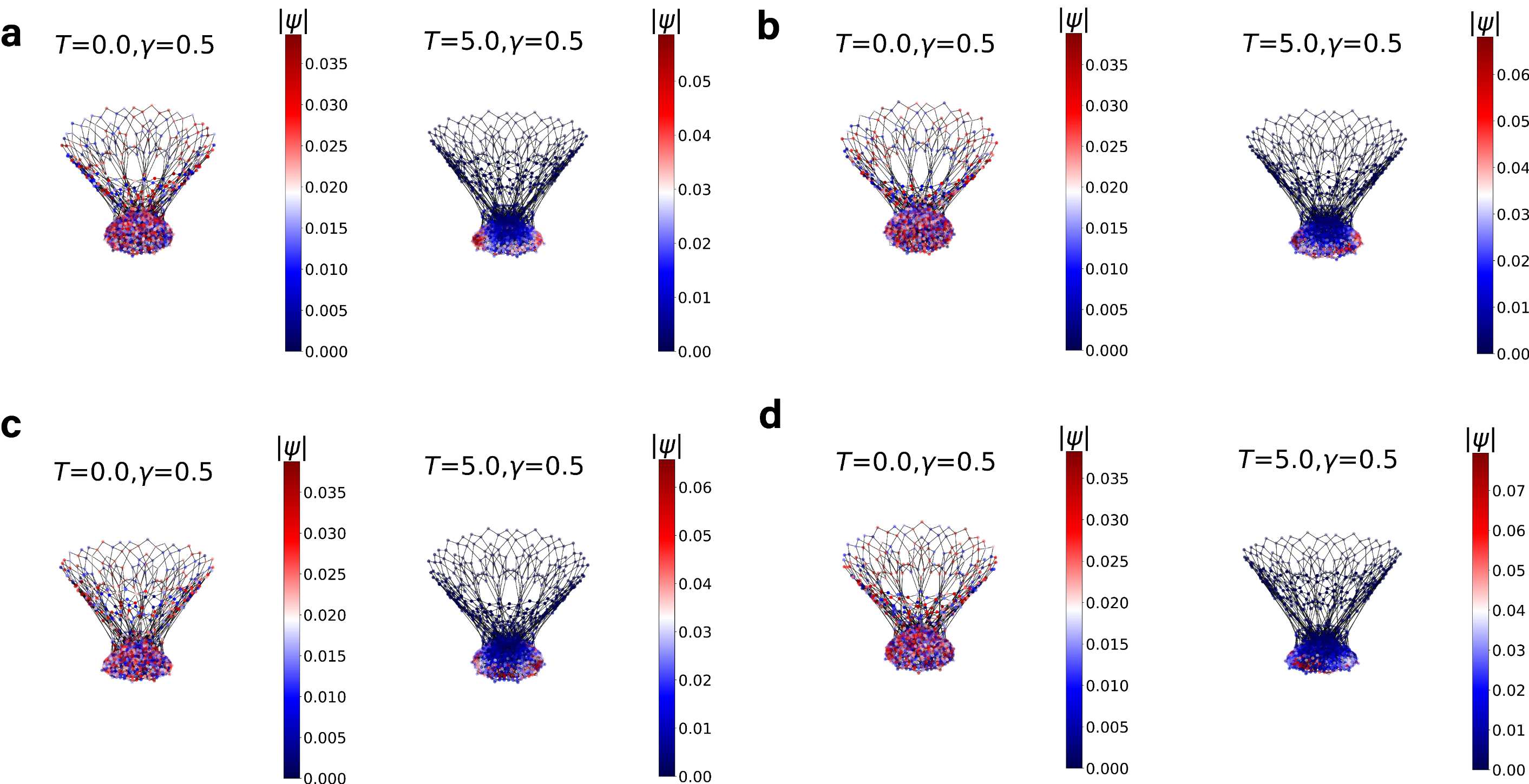}
	\caption{\textbf{a}-\textbf{d}: Dynamical evolution of four randomly prepared state in the Hilbert space graph for 5-boson sectors under PBCs, with lattice size $L=10$. We set $t_{1}=1.4, t_{2}=1.2$, and $\gamma=0.5$. The evolved state always eventually tend towards the localized skin cluster states shown in FIG.\ref{fig:suppfig13}}
	\label{fig:suppfig17}
\end{figure}
Moreover, in FIG.\ref{fig:suppfig17}, we prepare the different initial states in five-boson sector. All the evolutions approach the states with the skin localization at the bottom of the graph in FIG.\ref{fig:suppfig12}. Therefore, when we apply the quantum walk in the sector, the state will always meet the effective interface and exhibit the skin localization.
\newpage
\subsection{Scaling for N-boson sector}
In this section, we give the behavior of the scaling of a N-boson sector. Due to the correlated hopping in our model, in a N-boson Hilbert space, only a few states get involved. Thus, all the states involved generate one N-boson sector, and the  N-boson sector is embedded in the N-boson Hilbert space. For N-boson Hilbert space in dimension ${\rm Dim}_{Hilbert}$, only certain states are the computational basis for our model.

For one state $\left| \psi \right\rangle $ in the Hilbert space, if $\hat{H}\left| \psi \right\rangle\neq 0$, this state is included in our computational basis, and we counts all such non-trivial states to obtain the dimension of N-boson sector ${\rm Dim}_{Sector}$. One easy way to choose the states is to check the following local occupation:
\begin{equation}
	\begin{array}{r}
		\left\langle\psi \right|  \hat{n}_{i}\hat{n}_{i+3}\left| \psi \right\rangle > 0,	\\
		\left\langle\psi \right|   \hat{n}_{i}\left| \psi \right\rangle>1		 .\\
	\end{array}
\end{equation}

In the 3-boson Hilbert space, the state $\left|...10000100001... \right\rangle $ is one trivial state. For the following results on 4-boson and 5-boson cases, we find that ${\rm Dim}_{Sector}\approx{\rm Dim}_{Hilbert}$, and in the 5-boson case, the sector almost fills the Hilbert space.

\begin{figure}[h]
	
	\centering
	\includegraphics[width=0.99\linewidth]{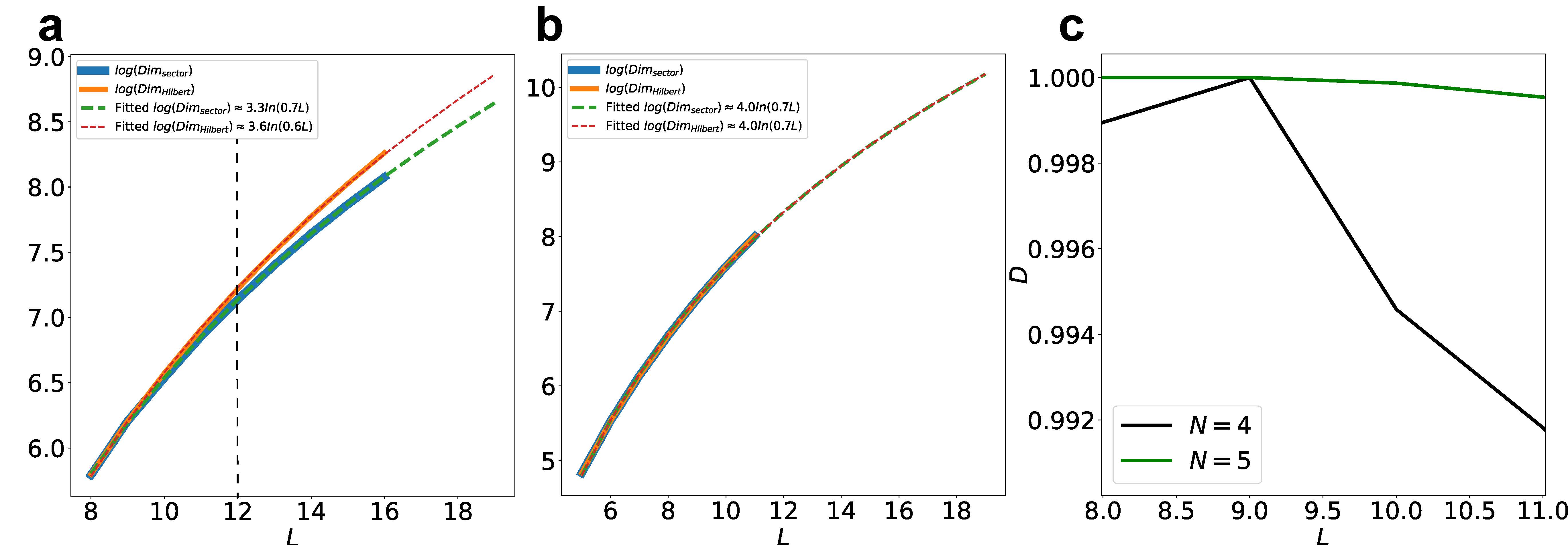}
	\caption{\textbf{a} and \textbf{b}: scaling of the dimensions $\log{\rm Dim}_{Sector}$  in the sector compared with the dimensions $\log{\rm Dim}_{Hilbert}$ for the whole Hilbert space with the size of the system $L$ for \textbf{a} 4-boson model and \textbf{b} 5-boson model. The dashed curves are the fitting results. For plot \textbf{b}, the five-boson sector grows in a similar scale with the whole Hilbert space and almost fills the full Hilbert space. \textbf{c} Fractal dimension $D=(\log{\rm Dim}_{Sector})/(\log{\rm Dim}_{Hilbert})\approx1$ for the sectors with filling $N=4$ and $N=5$ under the filling factor $\nu>\frac{1}{3}$. For the 4-boson case, the critical length is $L=12$ which is marked by the dashed line in \textbf{a}.}
	\label{fig:suppfig18}
\end{figure}

In our model, the interacting length is 3. Thus, once we have $3N>L$ ($\nu=\frac{1}{3}$ filling factor), the dimension of the sector is similar to the dimension of the full Hilbert space that ${\rm Dim}_{Sector}\approx{\rm Dim}_{Hilbert}=\binom{N+L-1}{L-1}$ under $\nu>\frac{1}{3}$.

Actually, the fractal dimension $D=(\log{\rm Dim}_{Sector})/(\log{\rm Dim}_{Hilbert})\approx1$ in FIG.\ref{fig:suppfig18} depends on the filling factor in our model. In FIG.\ref{fig:suppfig18} \textbf{a} and \textbf{b}, we compare two cases on the effect of the filling. For the case in FIG.\ref{fig:suppfig18} \textbf{a}, we consider the sector with 4 particles and critical length L=12, and we find the two curves fit well at the high filling and diverge around L=12 marked in the black dashed line. For another case in FIG.S20 \textbf{b}, we prepare sector with 5 particles under $L<15$, and consequently, we get ${\rm Dim}_{Sector}\approx{\rm Dim}_{Hilbert}$. For cases in FIG.\ref{fig:suppfig13} \textbf{b} and \textbf{c}, there are the skin localization in the finite fractal dimension under $\nu>1/3$.

%\textcolor{blue}{Any sector with n particles can be embedded in one larger Hilbert space with particles $N>n$. WLOG, we consider the ratio:
	%\begin{equation}\label{r}
	%	r(n,L)=\frac{\log{\rm Dim}_{sector,n}}{\log{\rm Dim}_{Hilbert,N=L}}.
	%\end{equation}}
	
	%\begin{figure}[h!]
	%	\centering
	%	\includegraphics[width=0.7\linewidth]{fig33}
	%	\caption{\textcolor{blue}{Ratio in Eq.\ref{r} for $n=4$ (red) and $n=5$ (blue) sectors with the dashed fitting curve as extension.}}
	%	\label{fig:fig33}
	%\end{figure}
	%\newpage
	
\begin{figure}[h!]
		\centering
		\includegraphics[width=0.9\linewidth]{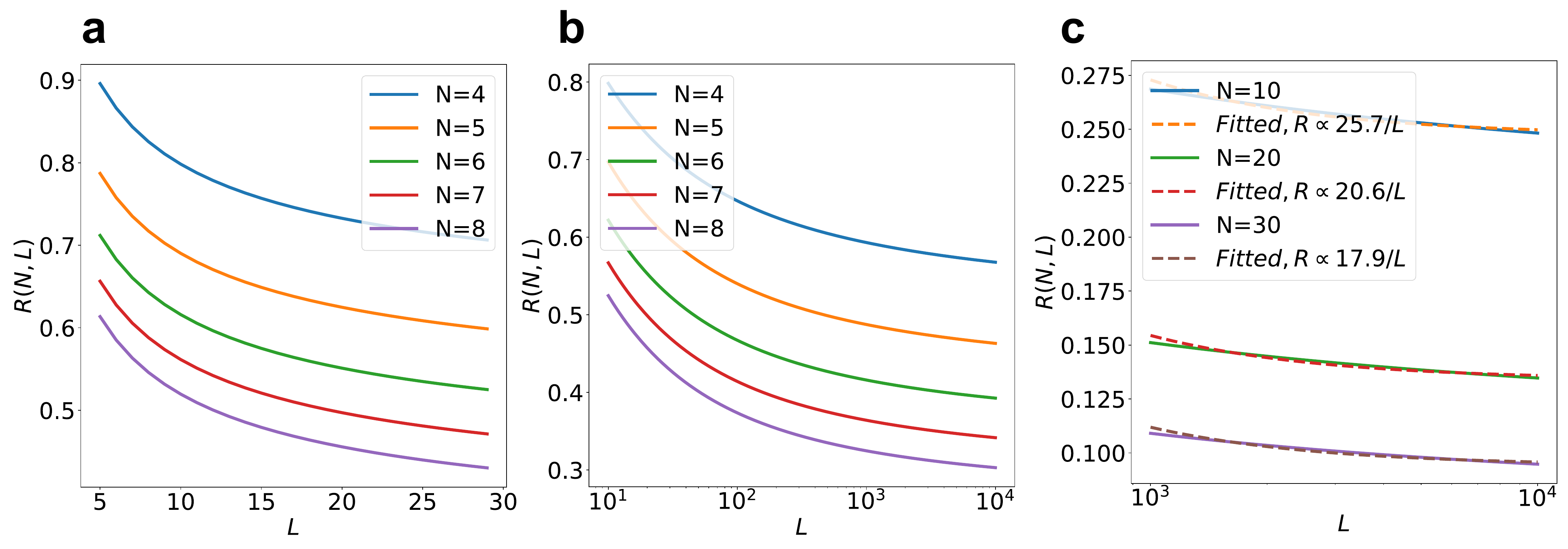}
		\caption{Ratio in Eq.~\eqref{RR} represented in different scales for the sector with 3 particles embedded in the different Hilbert spaces with N particles. In plot \textbf{c}, we apply the fitting (dashed curve) in the scale $R\propto\frac{1}{L}$ from Eq.~\eqref{scale} under the limit $L\gg N$. The non-zero ratio indicates the sector with 3 particles can still impact the whole system for reasonably large systems.}
		\label{fig:suppfig19}
	\end{figure}

	Since the graph of our model in the sector with 3 particle under PBCs can be mapped to one effective 2D lattice, we can easily compute the dimension as:
	\begin{equation}
		{\rm Dim}_{sector,n=3}=2L^{2}.
	\end{equation}
	Thus, for the sector with 3 particles, we define the ratio:
	\begin{equation}
		\label{RR}
		R(N, L)=\frac{\log \operatorname{Dim}_{\text {sector }, n=3}}{\log \operatorname{Dim}_{Hilbert}, N}=\frac{\log (2L^{2})}{\log \left(\left(\begin{array}{c}
				N+L-1 \\
				L-1
			\end{array}\right)\right)},
	\end{equation}
	which describes the weight of a sector with 3 particles embedded in one large Hibert space with N particles. In FIG.\ref{fig:suppfig19} (a), we find the ratio decays slowly under the condition of low filling. To understand the scaling of $R(N, L)$, we can approximate $R(N, L)$ as
	\begin{equation}
		\label{scale}
		R(N, L)\propto\frac{2\log (L)}{(N+L)\log(N+L)-(N)\log(N)-(L)\log(L)}\propto\frac{1}{L},
	\end{equation}
	which should remain finite in the limit with $L\gg N$, as long as $L$ is finite, as indicated in FIG.\ref{fig:suppfig19} (b) and (c). We emphasize that regardless of whether the fractal dimension vanishes or now, due to the robustness and non-local nature of the non-Hermitian effect, generic initial states always evolve towards the skin cluster states, as demonstrated in the previous subsection on dynamical evolution.

\end{document}